\documentclass[footinbib,twocolumn,amsmath,amstex,amssymb,mathfonts,superscriptaddress,prl]{revtex4}
\usepackage{graphicx}
\usepackage{color}
\usepackage{bm}
\usepackage{verbatim}

\usepackage{amsmath}
\usepackage{amssymb}
\usepackage{amsthm}
\usepackage{amsfonts}
\usepackage[caption=false]{subfig}

\def\blue{\textcolor{black}}

\newcommand{\be}{\begin{equation}}
\newcommand{\ee}{\end{equation}}

\newcommand{\la}{\langle}
\newcommand{\ra}{\rangle}

\renewcommand{\Im}{{\rm Im}\,}

\renewcommand{\vec}[1]{{\bf #1}}

\begin{document}
\title{Quenched topological boundary modes can persist in a trivial system}
\author{Ching Hua Lee}
\email{phylch@nus.edu.sg}
%\affiliation{Institute of High Performance Computing, A*STAR, Singapore, 138632}
\affiliation{Department of Physics, National University of Singapore, Singapore 117551}
\author{Justin C.W. Song}
\email{justinsong@ntu.edu.sg}
%\affiliation{Institute of High Performance Computing, A*STAR, Singapore, 138632}
\affiliation{Division of Physics and Applied Physics, Nanyang Technological University, Singapore 637371}
\date{\today}
\begin{abstract}
Topological boundary modes can occur at the spatial interface between a topological and gapped trivial phase and exhibit a wavefunction that exponentially decays in the gap. Here we argue that this intuition fails for a {\it temporal} boundary between a prequench topological phase that possess \blue{topological boundary} eigenstates and a postquench gapped trivial phase that does not possess any eigenstates in its gap. In particular, we find that characteristics of states (e.g., probability density) prepared in a topologically non-trivial system can persist long after it is quenched into a gapped trivial phase with spatial profiles that appear frozen over long times postquench. After this near-stationary window, \blue{topological boundary mode} profiles decay albeit, slowly in a power-law fashion. This behavior highlights the unusual features of nonequilibrium protocols enabling quenches to extend and control \blue{localized} states \blue{of both topological and non-topological origins}. 
\end{abstract}

\maketitle 
\section{Introduction}

Eigenstates dominate the long-lived excitations of a quantum system and determine its response. Recently, out-of-equilibrium protocols~\cite{Basov2017,Yao2007,OkaAoki2009,Lindner2011,Duan2006,jiang2011majorana,liu2013floquet,matthew2014floquet,zheng2014floquet,caio2015,potirniche2017floquet,li2018realistic,rudner2019} have emerged as a powerful means of controlling the states accessible in solid-state~\cite{Gedik2013,McIver2020,Mitrano2016,yap2018photoinduced} and cold-atomic optical lattice~\cite{Atala2013,Jotzu2014,Stuhl2015,goldman,flaschner2016experimental} platforms. For e.g., intense periodic drives can warp a topologically trivial electronic bandstructure to sustain states with non-trivial properties such as a finite Berry flux~\cite{Yao2007,OkaAoki2009,McIver2020,Atala2013,Jotzu2014,flaschner2016experimental} and induce spatially localized topological boundary modes (TBM) that traverse a bulk gap~\cite{torres,Stuhl2015,goldman}. Such out-of-equilibrium states are maintained only when the periodic drive is turned on; they cease to be eigenstates once the drive is switched off and effective hamiltonian quenches to a trivial phase~\cite{rigol2015}. 

\begin{figure}[ht!]
\label{fig:setup}
\includegraphics[scale=0.218]{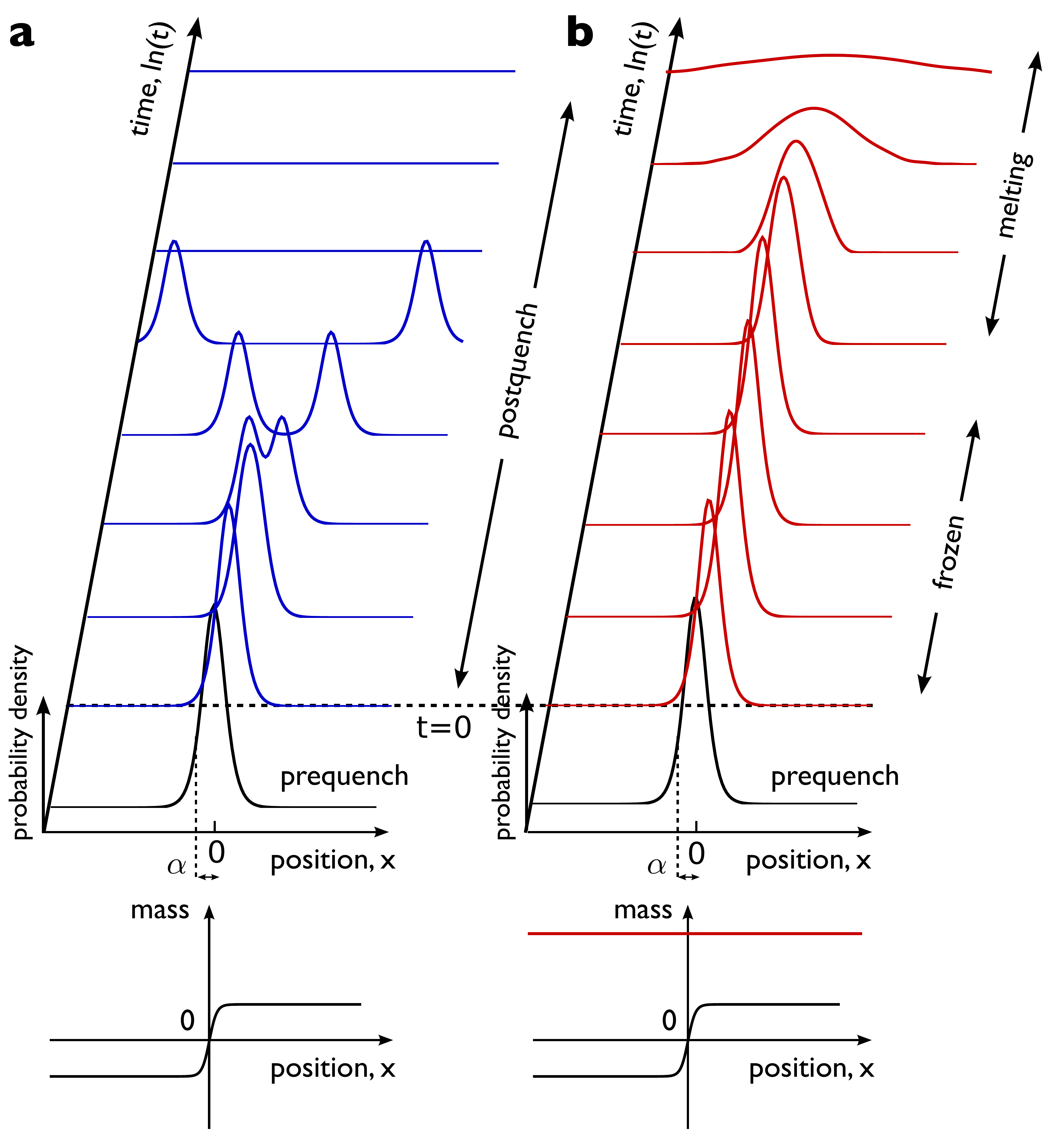}
\caption{\textbf{Emergence of frozen and melting regimes after a boundary mode is quenched. } 
{\bf a.} Postquench evolution of the probability density (PD) of a topological boundary mode (TBM) initially localized at the mass domain wall at position $x=0$; in this case where $m=0$ after the quench at time $t=0$, wavetrains move away from the domain wall for $t>0$. 
{\bf b.} Slow decay of the same TBM PD for large postquench uniform dimensionless mass $m\alpha/\hbar v=20$, $\alpha$ the domain wall width, $\hbar$ the Planck constant and $v$ the Fermi velocity. It displays a frozen PD regime as well as a melting regime exhibiting a power-law decay of the TBM PD. The pre-quench mass distribution is the same as in {\bf a}, with a domain wall at $x=0$, but the postquench mass $m\alpha/\hbar v = 20$ is large (red). In all panels PD snapshots are plotted using Eq.~(\ref{PD0}) taken at logarithmic time spacings. We used the initial TBM profile $\Phi(x)=\text{sech}(x/\alpha)$ and $\mathcal{N}=(\pi\sqrt{8\alpha})^{-1}$ from Eq.~(\ref{JR}).}
\label{Fig1}
\end{figure}

Here we argue that vestiges of spatially localized TBMs initially prepared in a gapped topologically non-trivial system persist long after it is quenched into a gapped trivial and uniform phase, see Fig.~\ref{Fig1}b. In particular, we find that even though TBMs are not eigenstates of the postquench phase and exist inside the postquench bulk gap, 
characteristics such as a spatially peaked TBM probability density (PD) persist and appear frozen over a long and tunable time window postquench. \blue{As we explain below, this frozen regime  %of slow-than-power-law decay 
is a consequence of the curvature of the gapped post-quenched dispersion profile, and applies to generic localized initial states. }
% even outside of the topological context\cite{caio2015}.}

This near-stationary nature of the TBM postquench PD (Fig.~\ref{Fig1}b, Fig.~\ref{Fig2}b) is particularly surprising given the rapid evolution of the postquench wavefunction. Indeed, this contrasts with recurrent Loschmidt-echo type responses found in many quenched systems~\cite{Altman2002,Levitov2008,Heyl2013} wherein observables oscillate as a function of time. 
Further, even as translational symmetry in the postquench hamiltonian is maintained, the postquench PD remains spatially localized (Fig.~\ref{Fig1}b) retaining its prequench profile. It decays as a power law at very long times. 

As we explain below, the persistence of PD arises from an interplay of TBM spatial localization and its fast dynamical evolution postquench. It is a striking example of how quantum systems pushed far out-of-equilibrium can possess very unusual properties with no analogue in equilibrium~\cite{Rudner2013,Wilson2016,caio2015,lee2018floquet}. We anticipate TBM persistence can be readily accessed in quenched systems that are directly sensitive to the PD distribution that include in ultracold atoms setups~\cite{Jotzu2014,Stuhl2015,flaschner2016experimental,alba2011seeing,hauke2014tomography,li2016bloch}, 
as well as in photonic~\cite{Rechstman2013} or circuit realizations~\cite{lu2019probing,helbig2019observation,hofmann2019reciprocal,bao2019topoelectrical}.

\section{Results}
{\it Probability density and quench protocol -- } The persistence of TBMs can be most easily illustrated by a time-dependent Dirac Hamiltonian (two-band) that undergoes a  
quench at time $t=0$ (Fig.~\ref{Fig1}):
\begin{eqnarray}
\hat{\vec{H}} 
(\vec p ,\vec r,t) = \hbar v \vec p \cdot \boldsymbol{\sigma} +\mathcal{M}(x,t)\sigma_3 = \bold d(\bold p, t)\cdot \boldsymbol{\sigma},
\label{}
\end{eqnarray}
where $v$ is the Dirac velocity, $\vec r=(x,y)$, $\vec p = (p_x, p_y)$, $\sigma_i$, $i=1,2,3$ are Pauli matrices, and 
time-varying mass
\begin{eqnarray}
\mathcal{M}(x,t)=\left\{\begin{matrix}
\,M(x) & \qquad t<0\\
\,m & \qquad t\geq 0
\end{matrix}
\right. 
\label{mass}
\end{eqnarray}
In the following, we shall refer to the pre- (post-) quench Hamiltonian as $\hat{\vec H}_i$ ($\hat{\vec H}_f$).  For $t<0$ before the quench, we have $M(x)$ satisfying $M(x<0)<0$ and $M(x>0)>0$, such that it describes a mass domain wall along the y-axis ($x=0$). Due to the jump in $\text{sgn}\,M(x)$, the domain wall carries a unit topological charge, and thus supports gapless topological boundary modes (TBM) $| \psi_{p_y}\ra$ that linearly disperse as $v\hbar p_y$. Their spatial wavefunction is well localized at the domain wall $x=0$: 
\begin{equation}
 \la \vec r | \psi_{p_y}\ra =\mathcal{N} e^{ip_yy} \, \Phi(x) \, | \Psi^{(0)} \ra , \quad | \Psi^{(0)} \ra = (1, i) /\sqrt{2},
\label{JR}
\end{equation}
where $\mathcal{N}$ is a normalization constant, and $\Phi(x)$ decays exponentially on both sides. 
The decay length, $\alpha$, defines the spatial extent of $\Phi(x)$. 

\begin{figure}[t!]
\label{fig:setup}
\includegraphics[scale=0.23]{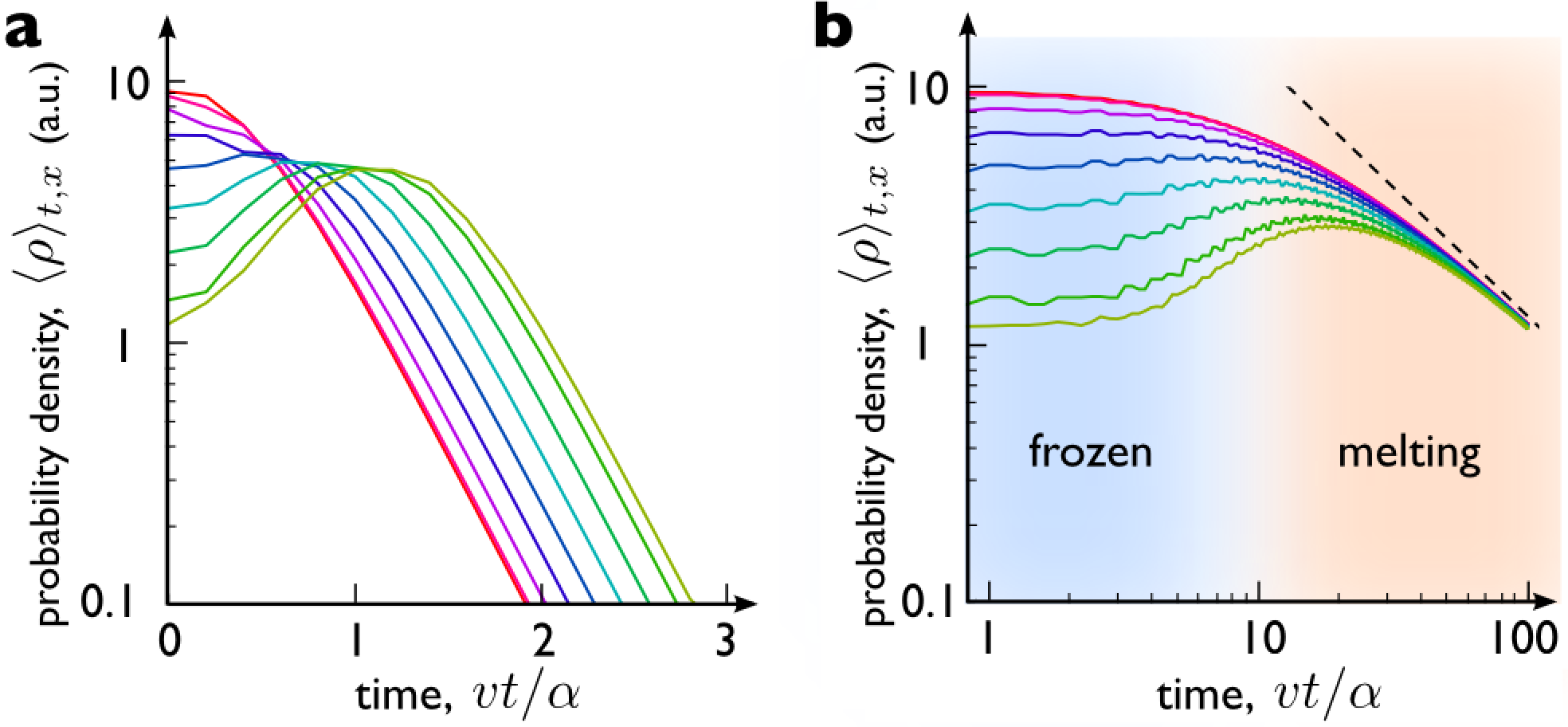}
\caption{\textbf{Frozen regime and subsequent power-law decay in the melting regime. } {\bf a.} Log-linear plot of dynamical evolution of postquench probability density $\langle \rho \rangle_{t,x}$ for small dimensionless mass $m \alpha/(\hbar v) = 0.1$, $\alpha$ the domain wall width, $\hbar$ the Planck constant and $v$ the Fermi velocity. This exhibits a fast exponential decay as displayed by the linear slope , with probability density attenuating rapidly. {\bf b.} Log-Log plot of postquench probability density $\langle \rho \rangle_{t,x}$ for large $m \alpha/(\hbar v) = 20$. In contrast to (a), this exhibits a very slow decay and has a evolution characterized by two distinct regimes: (blue shaded) a slow frozen regime where probability density hardly evolves, and (orange shaded) a power law melting regime where the probability density decays temporally as $t^{-1}$ (indicated by dashed line). For both panels, curves from top to bottom (red to green) indicate various positions away from the domain wall $x/\alpha=0 \to 1.2$ in steps of $0.15$. The probability density was obtained from direct numerical integration of Eq.~(\ref{PD0}) using the initial TBM profile $\Phi(x)=\text{sech}(x/\alpha)$ and $\mathcal{N}=(\pi\sqrt{8\alpha})^{-1}$ from Eq.~(\ref{JR}).}
\label{Fig2}
\end{figure}

Here we will consider a TBM state $| \psi_{p_y}\ra$ {\it prepared} in a prequench bulk gap with $|p_y|<\alpha^{-1}$. At $t=0$, the system is quenched via Eq.~(\ref{mass}) and the mass parameter becomes uniform in space. As a result, $|\psi_{p_y}\ra$ no longer exist as eigenstates of the postquench Hamiltonian and evolve as ${\rm exp}[{-i\hat{\vec H}_ft/\hbar}]| \psi_{p_y}\ra$. TBM $| \psi_{p_y}\ra$ postquench characteristics at $t>0$ are most saliently captured by its 
probability density (PD) 
\begin{equation}
\langle\rho\rangle_{t,x} = \la \psi_{p_y}| e^{i\hat{\vec H}_ft/\hbar} \hat{\vec{P}}_{\vec r}e^{-i\hat{\vec H}_ft/\hbar}| \psi_{p_y} \ra, \,\,\, \hat{\vec{P}}_{\vec r} = |{\vec r} \ra \la {\vec r} |,
\label{PD0}
\end{equation} 
with explicit $p_y$ dependence suppressed hereafter for brevity. Postquench, the initial boundary-localized TBM state is no longer an eigenstate of the new spatial-translation invariant Hamiltonian $\hat{\vec H}_f$; instead its temporal evolution can be understood from the Larmor precession of its projected components in the postquench eigenbasis of $\hat{\vec H}_f$. Due to the tightly localized (in $x$) profile of the initial TBM, many postquench $p_x$ eigenstates are accessed. As a result, their destructive interference and multi-frequency Larmor precession can lead to dephasing~\cite{Wilson2016} and decay of the initial TBM. Indeed, for $m \alpha/(\hbar v) \lesssim 1 $, we find through direct numerical integration of Eq.~(\ref{PD0}), the PD generically decays exponentially in time, Fig.~\ref{Fig2}{\bf a}.

However, when $m \alpha/(\hbar v) \gg 1$, we find that the PD no longer exhibits fast decay. Direct numerical integration of Eq.~(\ref{PD0}) yields an 
unusual regime wherein the initial TBM state \emph{freezes} maintaining its original localized (in $x$) profile (Fig.~\ref{Fig1}b and Fig.~\ref{Fig2}b) at short times. As we explain below, this frozen TBM stays locked over times $0 \leq t \ll \tau m\alpha/\hbar v$ controlled by the size of the postquench gap $m$; here the characteristic time $\tau = \alpha/2v$. At longer times, the TBM melts, displaying a much slower power-law decay. As we now explain, this arises from a near lock-step interference process between $p_x$ components enforced in the $m \alpha/(\hbar v) \gg 1$ regime.

To more concretely understand the freezing and subsequent melting of the TBM mode, we analyze 
the solution of $\langle\rho\rangle_{t,x}$ obtained through direct evaluation of Eq.~(\ref{PD0}):
\begin{equation}
\langle \rho\rangle_{t,x}=|C|^2+|\bold S|^2 -2\text{Im}[\bold S^*C]\cdot\hat{\vec{s}}-2(\text{Re}[\bold S]\times \text{Im}[\bold S])\cdot \hat{\vec{s}}, 
\label{PD1m}
\end{equation}
where $\vec S (x, t) = S_x \hat{\vec{x}} + S_y \hat{\vec{y}} + S_z \hat{\vec{z}}$, and $\hat{\vec s} = \la \Psi^{(0)} | \boldsymbol{\sigma} | \Psi^{(0)} \ra$ The quantities $C(x,t)$ and $\vec S (x, t)$ represent the even and odd temporal components of the postquench dynamical evolution of the PD and can be written as $C(x,t) = [(F_0 (x,t) + F_0 (x,-t)]/2$ and $S_{x,y,z} = [F_{x,y,z} (x,t) - F_{x,y,z} (x,-t)]/2i$ with  
\be
F_j(x,t) = \frac{\mathcal{N}}{2\pi}\int \bar\Phi(p_x)e^{ip_xx}e^{ i\varepsilon_pt/\hbar}f_j(p_x)dp_x, 
\label{eq:F}
\ee
where $j=0,x,y,z$, and the $j$-dependent weights as $f_0 =1$ and $(f_x,f_y,f_z) = \vec d(\vec p) /|\vec d(\vec p)|$. Here 
$\varepsilon_p=\sqrt{\hbar^2 v^2|\bold p|^2+m^2}$ is the (postquench) Dirac eigenenergy and $\bar{\Phi}(p_x)$ is the Fourier transform of TBM profile $\Phi(x)$. In obtaining Eq.~(\ref{PD1m}), we expanded ${\rm exp}[{-i\hat{\vec H}_ft/\hbar}]$ and $ \hat{\vec{P}}_{\vec r}$ in Eq.~(\ref{PD0}) as superpositions of the postquench energy eigenstates, and applied the Pauli matrix identity $\sigma_i\sigma_j=\delta_{ij}+i\epsilon_{ijk}\sigma_k$ repeatedly, see full details in Supplementary Information, {\bf SI}.

Initially at $t=0$, $\bold S (x,t=0) =0$ and $C (x,t=0) $ reduces to the spatial profile $\Phi(x)$ of the topological boundary state yielding Eq.~(\ref{PD1m}) that follows the profile of the TBM mode. For $t>0$, however, both the arguments of $\vec S(x,t)$ and $C(x,t)$ oscillate with time capturing the Larmor precession of the TBM projected components in the postquench eigenbasis. Indeed, given the wide range of $p_x$ eigenstates that the TBM state projects to, the temporal evolution of PD in Eq.~(\ref{PD1m}) involves (momentum-space) nonlocal interference between multiple distinct $p_x$ momentum modes, and can lead to a complex spatio-temporal evolution of the PD.

When $m \alpha/(\hbar v) \lesssim 1$, 
the dynamical phase factor ${\rm exp}[{ i\varepsilon_pt/\hbar}]$ in Eq.~(\ref{eq:F}) rapidly oscillates as a function of $p_x$. As a result, it suppresses the large momentum contributions to $C(x,t)$ (and $\bold S(x,t)$). Since the tight localization of the initial TBM depends on large $p_x$ momentum contributions, the initial state is rapidly eroded. On a physical level, this can be understood as an intrinsic dephasing phenomena where the multiple frequency 
oscillation get out of phase destroying the coherence of the initial state. In real space, this manifests as two wavetrains moving almost uniformly in opposite directions with a group velocity $v_g = v\hbar p_x/\sqrt{\Delta^2+\hbar^2 v^2p_x^2} \approx \pm v$ for small $\Delta^2=m^2+v^2 \hbar^2 p_y^2$ (see Fig.~\ref{Fig1}{\bf a})

{\it Frozen TBMs and Wick-rotated diffusion --} However, when $m \alpha/(\hbar v)  \gg 1$, the Larmor precession of the postquench state for various $p_x$ components oscillate in near lock-step, see below. Indeed, in this limit, the oscillation frequency $\varepsilon_p= \sqrt{\Delta^2+\hbar^2 v^2 p_x^2} \approx \Delta + (\hbar v p_x)^2/(2\Delta)$ is dominated by a large $\Delta= (m^2+\hbar^2 v^2 p_y^2)^{1/2}\approx m$. In the latter, we focus on small $p_y < \alpha^{-1}\ll m$. As the tight localization of the TBM state arises from large $p_x \sim \alpha^{-1}$ components, when $m \alpha/(\hbar v)  \gg 1$, the differences in the frequencies are severely suppressed. This near-lock-step oscillation dramatically slows the erosion of the TBM state and as we will now discuss, at short times can even halt it. To see this explicitly, we take $\varepsilon_p\approx \Delta + (\hbar v p_x)^2/(2\Delta)$ in Eq.~(\ref{eq:F}) and directly integrate:  
\be 
F_{0,z} (x,t) = \mathcal{N} \sqrt{\frac{i \Delta}{\hbar v^2 t}} e^{i\Delta t/\hbar} \Psi_t(x), 
\label{eq:gaussian}
\ee 
with the postquench profile 
\be
\Psi_t(x)=\int dx' \Phi(x+x') G_t(x'), \quad G_t(x) = e^{-i \Delta x^2/(2t \hbar v^2)}
\label{eq:Psit}
\ee
where $G_t(x)$ is an imaginary Gaussian kernel that results from the gapped 
dispersion in the large $m\alpha/(\hbar v)\gg 1$ limit, see {\bf SI}. 
This kernel $G_t(x)$ is the mathematical embodiment of the random interference processes between different momentum sectors, with its imaginary Gaussian nature paralleling Brownian motion scattering in a diffusive medium; it is fundamentally distinct from the dynamical behavior expected when the system is quenched into a gapless linearly hamiltonian, e.g., $m=0$ above. We note that $F_{x,y}$ contributions are small (suppressed in the $m \alpha/(\hbar v)  \gg 1$ limit) and do not contribute substantially to $\la \rho \ra_{t,x}$ profile~\cite{SuppMat}.  

For short times $0<t\ll m\alpha^2/(2\hbar v^2) = \tau m \alpha/ (\hbar v)$, 
the imaginary gaussian $G_t(x)$ in Eqs.~(\ref{eq:gaussian}) and (\ref{eq:Psit}) exhibit spatial oscillations far more rapid than $\Phi(x+x')$ varies in $x$. Here $\tau = \alpha/2v$ is an intrinsic  timescale. As a result, at these short times $F_{0,z} \approx \mathcal{N} \Phi(x) e^{i\Delta t/\hbar}$, so that $C(x,t) \approx \mathcal{N}\Phi(x) {\rm cos}(\Delta t/\hbar) $ and $S_{z} \approx \mathcal{N}\Phi(x) {\rm sin}(\Delta t/\hbar) $. %
Substituting into Eq.~(\ref{PD1m}) we obtain the PD: 
\be \langle \rho\rangle_{t,x} \approx\mathcal{N}^2 |\Phi(x)|^2 + \mathcal{O}\left(\frac{\hbar v}{m\alpha}\right), \quad 0<t< \frac{\tau m \alpha}{\hbar v}.
\label{eq:frozen}
\ee
We note PD in this regime is flat to infinite order in $vt/\alpha$, Fig.~\ref{Fig2}{\bf b} (blue shaded); any temporal decay in this frozen regime is non-analytic and {\it slower} than any power law.  As a result, we term this a region of {\it frozen} PD. This is particularly surprising since the prequench TBM state is not an eigenstate of the postquench Hamiltonian $\hat{\vec H}_f$. Perhaps even more striking is the fact that the period over which the TBM's PD is frozen can be controlled by the postquench gap size $m$; increasing $m$ gives a wider frozen window for TBM PD, Eq.~(\ref{eq:frozen}). This can be understood physically \blue{by noting that the time window over which the PD remains frozen can be recast in a more physical intuitive way as the ratio of the spatial confinement of the prequench domain wall, $\alpha$, and the typical group velocity of wavepackets postquench; larger $m$ translates to slower postquench group velocities and hence longer frozen time window.} 

At long times $t\gg m\alpha^2/(2\hbar v^2)$, 
the situation is dramatically different with the imaginary gaussian $G_t(x)$ in Eqs.~(\ref{eq:gaussian}) and (\ref{eq:Psit}) 
slowly varying over large ranges of $x$. As a result, the frozen states ``melts'' with its spatial profile spreading slowly out according to the convolution $\Psi_t(x)$ and an overall PD amplitude decaying as $t^{-1}$:
\be
\langle \rho\rangle_{t,x} \to \frac{ m}{2t \hbar v^2} \mathcal{N}^2|\Psi_t(x)|^2, \quad t \gg \frac{\tau m \alpha}{\hbar v}. 
\label{eq:melting}
\ee
This $t^{-1}$ power-law melting decay (dashed line) conforms with that found from direct numerical integration in Fig.~\ref{Fig2}{\bf b} (orange shaded), and even exhibits a data collapse for PD taken at different values of $x$. In this regime, the spatial extent/width of $|\Psi_t(x)|^2$ grows as $t$ thereby conserving the total probability density over all space. \blue{We note that the persistence of TBM postquench is consistent with the edge currents that found in quantum quenches of Chern insulators~\cite{caio2015}. }

To understand this long-time behavior more intuitively, we note that $G_t(x)$ can be interpreted as a 1D heat kernel corresponding to a Wick-rotated ``diffusion'' process (i.e. evolution by Schr\"{o}dinger's equation involving the $id/dt$ instead of $d/dt$ operator) with diffusion constant $-i\hbar/2\Delta$. 
While scattering processes in ordinary diffusion processes lead to a ``Gaussian-blurred'' distribution characterized by a (real) Gaussian decaying kernel, the imaginary Gaussian kernel $G_t(x)$ represents the rapid interference effects from multiple non-coherent contributions in Schr\"{o}dinger evolution.
A large $\Delta$ results in slow spreading of the initial state, which by Eq.~(\ref{eq:Psit}) is asymptotically governed by power-law decay, instead of Gaussian decay as in the more familiar real-time diffusion scattering processes. Physically, the slow melting TBM behavior can be likened to that of slow diffusion found in classical systems with large inertial masses; the large $m\alpha/(\hbar v)\gg 1$ limit corresponds to the regime where postquench modes are energetically inaccessible. 

{While we have concentrated on a continuum description of the frozen and melting regimes in the main text, both frozen and melting behavior persist even on a lattice model (see detailed numerics of quench dynamics on a lattice in {\bf SI}) as well as at hard-wall boundaries. This agreement arises because the frozen and melting regimes are most sensitive to the long-wavelength evolution of the TBM and the mismatch between the TBM confinement, $\alpha$, and the postquench Compton lengthscale $\hbar v/m$. Indeed, this mismatch and the ensuing quench dynamics we describe here can manifest for other initially localized modes, topological or otherwise.}

{\it Pseudo-spin precession and creep current -- } The frozen-in-time PD profile at short times described in Eq.~(\ref{eq:frozen}) hides the fact that the TBM modes are not eigenstates of the postquench Hamiltonian $\hat{\vec H}_f$. Are all other observables similarly frozen in time as well? To further interrogate the TBM postquench, we consider its local (spatially-resolved) pseudo-spin expectation 
\begin{equation}
\langle\boldsymbol{\sigma}\rangle_{t,x} = \la \psi_{p_y}| e^{i\hat{\vec H}_ft/\hbar} \hat{\vec{P}}_{\vec r} \boldsymbol{\sigma} \hat{\vec{P}}_{\vec r}e^{-i\hat{\vec H}_ft/\hbar}| \psi_{p_y} \ra, 
\label{eq:localspin}
\end{equation} 
with explicit $p_y$ dependence suppressed hereafter for notational brevity. $\langle\boldsymbol{\sigma}\rangle_{t,x}$ encodes the spinor-wavefunction information of the TBM. prequench, $\langle\boldsymbol{\sigma}\rangle$ is aligned along $\hat{\vec s}$ in the $\hat{\vec e}_2$ direction; here $\vec e_{1,2,3}$ denote the $\vec x, \vec y, \vec z$ directions in a Bloch sphere. However, postquench, the eigenstates of $\hat{\vec H}_f$ generically possess spinor components in all three-directions. As a result, the dynamical evolution of $\langle\boldsymbol{\sigma}\rangle_{t,x}$ postquench involves a complex intertwining of precession and interference between wave components composing the spatially localized profile of the TBM state. Indeed, $\langle\boldsymbol{\sigma}\rangle_{t,x}$ possesses a dynamics that generically departs from that of the Bloch equation~\cite{SuppMat}. 

To see this, we directly evaluate Eq.~(\ref{eq:localspin}) in the same fashion as Eq.~(\ref{PD1m}) above producing the closed form 
\begin{align}
\langle \boldsymbol{\sigma}\rangle_{t,x} 
&= \hat{\vec{s}} (|C|^2-|\bold S|^2) +2\text{Re}[\bold S^*C]\times \hat{\vec{s}} - 2\text{Im}[\bold S^*C] \nonumber \\ & 
+ 2\text{Re}[(\bold S^*\cdot\hat{\vec{s}} )\bold S]+2\text{Re}[\bold S]\times \text{Im}[\bold S]. 
\label{eq:localspindynamics}
\end{align}
The first term tracks the persistence of the initial pseudospin direction $\hat{\vec{s}}$, while the second term represents a ``pure'' precession contribution; the other terms correspond to additional dynamical contributions from $\bold S$, which arises physically from projection onto postquench eigenstates. Eq.~(\ref{eq:localspindynamics}) in fact describes the dynamical solution of \emph{any} two-component state with an initial spatially inhomogeneous profile, see {\bf SI}. In our case, Eq.~(\ref{eq:localspindynamics}) contains information on how the localization of them TBM interplays with precession effects. This generically yields a complex spatio-temporal and spin-dependent evolution, see explicit example in {\bf SI}.

\begin{figure}
\includegraphics[width=\columnwidth]{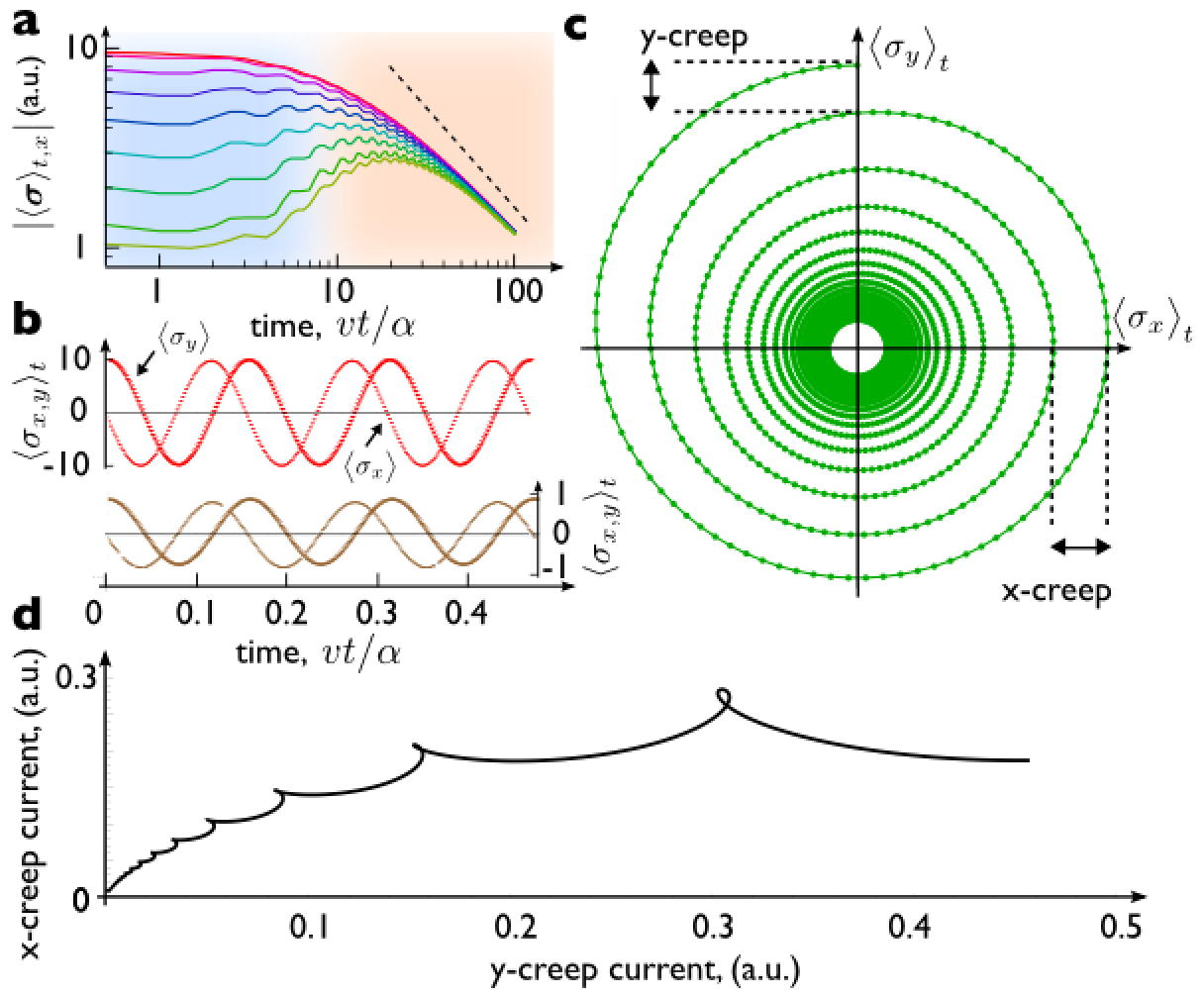}
\caption{\textbf{Evolution of pseudospin expectation and its creep after the quench.} {\bf a.} Log-log plot of dynamical evolution of the local pseudo-spin magnitude $|\langle\boldsymbol{\sigma}\rangle_{t,x}|$ for large dimensionless mass $m \alpha/(\hbar v) = 20$, $\alpha$ the domain wall width, $\hbar$ the Planck constant and $v$ the Fermi velocity. It exhibits a frozen (blue) and a melting regime (orange); dashed line indicates a $t^{-1}$ temporal power-law decay. The curves from top to bottom (red to green) indicate various positions away from the domain wall $x/\alpha=0\rightarrow 1.2$ in steps of $0.15$. {\bf b.} $\la\sigma_x\ra$ and $\la \sigma_y\ra$ oscillation at $x=0$ (red) and $x/\alpha=f1.2$ (brown). {\bf c.} Trajectory of  $\langle\boldsymbol{\sigma}\rangle_{t,x}$ for $x=0$ for $m \alpha/(\hbar v) = \pi $ exhibiting a decay in the pseudo-spin density over time. {\bf d} The decay over each cycle leads to a creep current: a non-vanishing drift current of the topological boundary mode. Creep current estimated by averaging $\la\sigma_{x,y}\ra_{t,x}$ over a period twice, \blue{ such as to smooth out fluctuations in the period. The small kinks represent subleading second-order corrections in the power-law decay. } }
\label{Fig3}
\end{figure}

However, when $m\alpha/\hbar v \gg 1$, the pseudo-spin expectation $\langle\boldsymbol{\sigma}\rangle_{t,x}$ of the TBM mode can act like a localized block spin with TBM pseudo-spin expectation at each $x$ precessing in unison. Similar to the PD described above, this also leads to a frozen regime, where the magnitude of the $|\langle\boldsymbol{\sigma}\rangle_{t,x}|$ pseudo-spin is preserved for a sizeable period of time, Fig.~\ref{Fig3}a. Intuitively, in this limit, the $\bold d=(v\hbar p_x,0,m)$ precession axis points strongly towards in the $\hat{\bold e}_3$ direction, causing the initial pseudospin $\hat{\vec{s}}$ to precess largely in the $\hat{\bold e}_1$-$\hat{\bold e}_2$ plane. Since the precession axis does not shift much across different $p_x$ and $p_y$ sectors due to large $m$, destructive interference is minimized and the quenched TBM states periodically return to their initial configurations: $|\langle\boldsymbol{\sigma}\rangle_{t,x}|$ is preserved. Indeed, in this regime, various positions in the TBM have $\la \sigma_{x,y}\ra_{x,t}$ that oscillate in phase Fig.~\ref{Fig3}b.
At longer times $t\gg \tau m \alpha/\hbar v$, $|\langle\boldsymbol{\sigma}\rangle_{t,x}|$ melts, and decays as a power-law $t^{-1}$, Fig.~\ref{Fig3}a, see also {\bf SI} where the power-law is derived.

Since $\hbar^{-1}\partial \hat{\rm H}_f/\partial \vec p = v (\sigma_x,\sigma_y)$, the pseudo-spin oscillations indicate a cyclotron-type motion of the TBM mode; in the large $m\alpha/\hbar v \gg 1$, the oscillations are largely locked to a frequency $2m/h$. Interestingly, when the magnitude of the pseudo-spin decays, the value of $\la \sigma_x \ra$ or $\la \sigma_y\ra$ does not come back to itself after a full revolution, Fig.~\ref{Fig3}c. This decay in pseudo-spin leads to a {\it creep} current: an uncompensated drift current of the TBM (see estimate in Fig.~\ref{Fig3}d) that drifts along the domain wall. 

The persistence of TBM characteristics can form the basis for a strategy to extend TBM features long after the topological system that supports it is gone: namely, by quenching to a large gap wherein postquench eigenstates are energetically inaccessible. Since the persistent TBM PD arises from the fast postquench Larmor precession enforced by large $m$, we expect that it is robust against inelastic scattering with energies far smaller than the gap scale as well as slowly varying disorder with typical lengths longer than the TBM width. The frozen and melting quenched PD regimes can be readily prepared and measured in ultra-cold atom optical lattices~\cite{Jotzu2014,Stuhl2015,flaschner2016experimental,alba2011seeing,hauke2014tomography} with its pseudospin components of the wavepacket independently extracted~\cite{alba2011seeing,hauke2014tomography}, and its real-space dynamics tracked~\cite{flaschner2016experimental}. 

{While we expect that the frozen and melting quench dynamics that we discuss here to be most easily realized in ultra-cold atomic optical lattices, or in photonic~\cite{Rechstman2013} or circuit platforms~\cite{lu2019probing,helbig2019observation,hofmann2019reciprocal,bao2019topoelectrical} wherein interactions are weak, it is interesting to consider how the quench dynamics will behave in electronic systems wherein interactions can be present. In an electronic platform, interactions can lead to dephasing that can disrupt the lock-step precession of the postquench states that underlie the Frozen and melting regime~\cite{Wilson2016}. As a result, we expect the unusual frozen and melting regimes to only manifest at times shorter than the dephasing time of the electronic system.}

Perhaps most striking, is the contrast in behavior of TBMs across a sharp {\it temporal} boundary (as realized in the quench at $t=0$) against those found at spatial boundaries. In the latter, boundary states (topological or otherwise) that exist in an energy gap typically exponentially decay over space into a gapped insulating bulk. Whereas in the former, TBM PD persists, freezing for a long window then decaying in a power-law fashion even when there are no eigenstates in trivial gap. This underscores the stark difference between sharp spatial and temporal interfaces and displays the dichotomy between equilibrium and out-of-equilibrium responses.

\vspace{2mm} 
{\it Acknowledgements} -- We acknowledge useful conversations with Mark Rudner and Yidong Chong. This work was supported by the Singapore National Research Foundation (NRF) under NRF fellowship award NRF-NRFF 2016-05, a Nanyang Technological University start-up grant (NTU-SUG), and a Singapore MOE Academic Research Fund Tier 3 Grant MOE 2018-T3-1-002. J.C.W.S acknowledges the hospitality of the Aspen Center for Physics, which is supported by National Science Foundation grant PHY-1607611, where part of this work was performed. 

\section{Methods}

\section{Postquench dynamical evolution}

In this section, we review how the topological boundary mode (TBM) in Eq.~(\ref{JR}) of the main text and its characteristics (namely probability density (PD) and pseudospom expectation) evolve postquench. For generality, we do not make any assumption on the form of the initial state (unless otherwise stated), other than it can be spatially inhomogeneous in the x-direction. We shall also assume the most general post-quench Hamiltonian that is translation-invariant in both directions. Hence our following results are applicable for a broader class of quench settings than those discussed in the main text.

\subsection{Wavefunction evolution}

The postquench ($t>0$) system is described by the Hamiltonian $\hat{\vec H}_f(\bold p)=\bold d(\bold p)\cdot \boldsymbol{\sigma}$ with bulk eigenstates $|\varepsilon_{\bold p}^\pm \ra$ with eigenenergies $\pm \varepsilon_p = \pm |\vec d(\vec p)|= \pm \sqrt{\hbar^2 v^2|\bold p|^2+m^2}$. 
The wavefunction of the TBM initial state [see Eq.~(\ref{JR}) of the main text] at $t=0$ can be written as $ \la \vec r| \psi_{p_y}\ra 
=\sum_{p_x}\mathcal{N}\bar\Phi(p_x)e^{ip_xx}e^{ip_yy} | \Psi^{(0)} \ra $, where $\bar\Phi(p_x)$ is the Fourier transform of $\Phi(x)$. The time-evolved wavefunction postquench can be directly evaluated as $ \psi_{p_y} (\vec r, t) = \la \vec r| e^{-i\hat{\vec H}_ft} | \psi_{p_y}\ra$. 

Writing ${\rm exp}(-i\hat{\vec H}_ft)  = \sum_{\vec p, \pm}  e^{\mp i\varepsilon_pt} | \varepsilon_{\bold p}^\pm\ra \la\varepsilon_{\bold p}^\pm |$, and noting that $ | \varepsilon_{\bold p}^\pm\ra \la\varepsilon_{\bold p}^\pm | = \tfrac{1}{2}\big(\mathbb{I}\pm \hat{\vec d}(\bold p)\cdot \boldsymbol{\sigma}\big)|\bold p\rangle\langle \bold p|$, where $\la \vec r | \vec p \ra = e^{i\vec p\cdot \vec r}$ is the wavefunction of the momentum eigenstate, and $\hat{\vec d}(\vec p) = \vec d(\vec p) /|\vec d(\vec p)|$, we obtain
\be
\psi_{p_y} (\vec r, t) = \sum_{p_x,\bold r'} e^{i\bold p\cdot(\bold r-\bold r')}(\cos \varepsilon_pt-i \hat{\vec d}(\bold p)\cdot \boldsymbol{\sigma} \sin\varepsilon_pt) \la \vec r' |\psi_{p_y} \ra 
\label{eq:Sevolution-intermediate}
\ee
In obtaining Eq.~(\ref{eq:Sevolution-intermediate}) we summed across the $\pm$ bands as well as inserted the resolution of the identity $\sum_{\vec r'} |\vec r' \ra \la \vec r'|$. Summing across $\vec r', p_x'$ by writing out $ \la \vec r| \psi_{p_y}\ra  =\sum_{p_x}\mathcal{N}\bar\Phi(p_x)e^{ip_xx}e^{ip_yy} | \Psi^{(0)}\ra$ we obtain the compact form
\be
\psi_{p_y}(\vec r, t) = e^{ip_y y} \Big[ C(x,t) - i \vec S(x,t) \cdot \boldsymbol{\sigma} \Big] | \Psi^{(0)} \ra
\label{eq:Sevolution}
\ee
where $C$ and $\vec S$ are the same as the main text. We reproduce these here for the convenience of the reader 
\begin{align}
C(x,t) &= \mathcal{N} \sum_{p_x} \bar\Phi(p_x)e^{ip_xx} \cos \varepsilon_pt  \\ 
\vec S (x,t) &= \mathcal{N} \sum_{p_x} \hat{\vec d}(\bold p) \bar\Phi(p_x)e^{ip_xx} \sin \varepsilon_pt  
\end{align} 
In Eq.~(\ref{eq:Sevolution}), we see that $\psi(t,\bold r)$ comprises two pseudospinor contributions, one proportional to $\cos\varepsilon_pt$ in the direction of the original spinor $|\Psi^{(0)}\rangle$, and another proportional to $i\sin\varepsilon_pt$ for which $|\Psi^{(0)}\rangle$ has undergone a $\hat{\vec d}(\bold p)\cdot\boldsymbol{\sigma}$ spinor rotation.

\subsection{Probability density dynamics}

The evolution of the probability density postquench can be obtained from direct evaluation of Eq.~(\ref{PD0}) of the main text. This amounts to finding the square amplitude of the wavefunction above, $|\psi_{p_y} (\vec r, t)|^2$. This can be evaluated as
\be
\la \rho\ra_{t,x} =  \la  \Psi^{(0)} |  \Big[ C^* + i \vec S^*\cdot \boldsymbol{\sigma} \Big]   \Big[ C- i \vec S \cdot \boldsymbol{\sigma} \Big]  | \Psi^{(0)} \ra,
\ee
where we have suppressed the explicit $x,t$ dependence of $C, \vec S$ for brevity. This expression can be readily simplified by recalling the vector identity:
\be
(\vec a \cdot \boldsymbol{\sigma}) (\vec b \cdot \boldsymbol{\sigma}) = \vec a \cdot \vec b \mathbb{I} + i (\vec a \times \vec b) \cdot \boldsymbol{\sigma}.
\ee
This yields the compact expression for PD as 
\be
\la \rho\ra_{x,t} = |C|^2 + |\vec S|^2 + i(\vec S^* C - C^* \vec S) \cdot \hat{\vec s} + i (\vec S^* \times \vec S) \cdot \hat{\vec s}
\label{eq:S-rho}
\ee
where $  \hat{\vec s} = \la  \Psi^{(0)} | \boldsymbol{\sigma} |  \Psi^{(0)} \ra$. Writing $\vec S^* C - C^* \vec S = 2i \Im (\vec S^* C)$ we obtain Eq.~(\ref{PD1m}) in the main text. 

We note, parenthetically, when $\vec d(\vec p)$ describes a Dirac model, further simplifications can be made to Eq.~(\ref{eq:S-rho}). For example, for Dirac models $\varepsilon_p$ is even in $p_x$. This forces $C$ to be real, and $\text{Re}[\bold S]$ ($\text{Im}[\bold S]$) to arise solely from the even(odd) components of $\hat{\vec{d}}(p_x,p_y)$. Hence $\text{Im}[\bold S]\parallel \hat{\bold e}_1$ and $\text{Re}[\bold S]\perp \hat{\bold e}_1$, and the time-evolved probability density reads as
\begin{equation}
\langle \rho\rangle_{x,t}=C^2+(\text{Im}[\bold S]\cdot \hat{\bold e}_1-\text{Re}[\bold S]\hat {\bold e}_3)^2+(\bold S\cdot \hat{\bold e}_2)^2,
\label{PD2}
\end{equation}
with the last term nonzero only for $p_y\neq 0$.

\subsection{Pseudospin dynamics}

The evolution of the pseudospin expectation (density) postquench can be evaluated in the same fashion as above. Using the $C,\vec S$ notation above, the pseudospin expectation density in Eq.~(\ref{eq:localspin}) of the main text can be re-written compactly as
\be
\la \boldsymbol{\sigma}\ra_{t,x} =  \la  \Psi^{(0)} |  \Big[ C^* + i \vec S^*\cdot \boldsymbol{\sigma} \Big]   \boldsymbol{\sigma} \Big[ C- i \vec S \cdot \boldsymbol{\sigma} \Big]  | \Psi^{(0)} \ra,
\ee
This expression can be readily simplified by repeated application of the identiy $\sigma_a \sigma_b = \delta_{ab} \mathbb{I}  + i\epsilon_{abc} \sigma_c$ and recalling the identity $\epsilon_{abc}\epsilon_{ade} = \delta_{bd}\delta_{ce} - \delta_{be}\delta_{cd}$. Here $\epsilon_{abc}$ is the Levita-Cevita symbol, and $a,b,c,d,e$ indices run over $x,y,z$.  Focussing on the $a$-th component of the pseudospin expectation density, we obtain
\begin{align}
[\la \boldsymbol{\sigma}\ra_{t,x} ]_a &= |C|^2 \hat{\vec s}_a - 2 \Im(\vec S^*_a C) + \epsilon_{abc} \big( C^* \vec S_b \hat{\vec s}_c + C \vec S^*_b \hat{\vec s}_c \big) \nonumber \\
&+ \vec S^*_a \vec S_b \hat{\vec s}_b + \vec S^*_b \hat{\vec s}_b \vec S_a - \vec S_b^* \vec S_b \hat{\vec s}_a +  i \epsilon_{abc} \vec S_b \vec S^*_c
\label{sigmaxt}
\end{align}
Re-writing in terms of real and imaginary parts of $\vec S$ and $C$, we obtain Eq.~(\ref{eq:localspindynamics}) in the main text. In obtaining Eq.~(\ref{eq:localspindynamics}) we have noted the identity $ (\vec u+i\vec v) \times (\vec u - i \vec v) = -2i \vec u \times \vec v$, where $\vec u$ and $\vec v$ are vectors with real components.  

Similar to that discussed above, simplifications to Eq.~(\ref{sigmaxt}) arise when $\varepsilon_p$ is 
even in $p_x$. This forces $C$ to be real. In that case, $\text{Re}[\bold S]$ ($\text{Im}[\bold S]$) arises from the integral of the even(odd) components of $\hat{\vec{d}}(p_x,p_y)$.
If $\hat{\vec d}(\bold p)$ is furthermore purely even/odd, Eq.~(\ref{sigmaxt}) simplifies to
\begin{equation}
\la \boldsymbol{\sigma}\ra_{t,x}^\text{even}=\hat{\vec{s}}(|C|^2-|\bold S|^2) +2\text{Re}[\bold S^*C]\times \hat{\vec{s}}+2\text{Re}[(\bold S^*\cdot\hat{\vec{s}})\bold S]
\label{sigmaeven}
\end{equation}
and
\begin{equation}
\la \boldsymbol{\sigma}\ra_{t,x}^\text{odd}=\hat{\vec{s}}(|C|^2-|\bold S|^2)-2\text{Im}[\bold S^*C]+2\text{Re}[(\bold S^*\cdot\hat{\vec{s}})\bold S].
\label{odd}
\end{equation}
In either case, the last term of Eq.~(\ref{sigmaxt}) always disappears, since it requires $\bold S$ to have both real and imaginary parts.

Recalling that the TBM modes discussed in the main text have $\hat{\vec{s}}=\hat{\bold e}_2$, one also obtains the squared expected pseudospin magnitude (not to be confused with $\langle \rho\rangle_{t,\bold r}$)
\begin{equation}
|\la \boldsymbol{\sigma}\ra_{t,x}|^2=(|C|^2+|\bold S|^2-2(\bold S\cdot e_1)(\bold S\cdot e_3))^2
\label{sigmaxtmag}
\end{equation}
From the rotational invariance of Eq.~(\ref{sigmaxt}), one can also easily deduce the expressions valid for general initial states with $\hat{\vec{s}}\neq \hat{\bold e}_2$.

\section{Frozen and melting regimes} 

In this section, we provide a detailed description of how both frozen and melting regimes arise in the limit of large $m\alpha/\hbar v\gg 1$. Before we proceed, we note $C$ and $\vec S$ can be expressed as $C(x,t) = [(F_0 (x,t) + F_0 (x,-t)]/2$ and $S_{x,y,z} = [F_{x,y,z} (x,t) - F_{x,y,z} (x,-t)]/2i$ with  
\be
F_j(x,t) = \mathcal{N}\int \bar\Phi(p_x)e^{ip_xx}e^{ i\varepsilon_pt/\hbar}f_j(p_x)\frac{dp_x}{2\pi}, 
\label{eq:S-F0}
\ee
where $j=0,x,y,z$, and the $j$-dependent weights as $f_0 =1$ and $(f_x,f_y,f_z) = \vec d(\vec p) /|\vec d(\vec p)|$. Since the dynamics of $C(x,t)$ and $\vec S(x,t)$ are controlled by $F_j(x,t)$, we will analyze the temporal dynamics of $F_j (x,t)$ in the limit of $m\alpha/\hbar v\gg 1$. 

In so doing, we first note that in the large postquench mass limit we can write 
\be
\varepsilon_p= \sqrt{\Delta^2+\hbar^2 v^2 p_x^2} \approx \Delta + (\hbar v p_x)^2/(2\Delta)
\label{eq:S-energyapprox}
\ee
where $\Delta= (m^2+\hbar^2 v^2 p_y^2)^{1/2}$. We note parenthetically, when the postquench mass is large and by focussing on small $p_y < \alpha^{-1}\ll m/\hbar v$, we have $\Delta \approx m$.

The  tight localization of the TBM state arises from large $p_x \sim \alpha^{-1}$ components. When $m \alpha/(\hbar v)  \gg 1$, the differences in their precession frequency are very much suppressed and can be well approximated by Eq.~(\ref{eq:S-energyapprox}). Applying Eq.~(\ref{eq:S-energyapprox}) into the precession frequency of Eq.~(\ref{eq:S-F0}) we obtain 
\be
F_j(x,t)\approx  \frac{\mathcal{N}e^{ i\Delta t/\hbar}}{2\pi} \int \Phi(x') e^{ip_x(x-x')}e^{ i\frac{\hbar(vp_x)^2t}{2\Delta}}f_j dp_x dx', % \notag\\
\label{eq:S-Fj}
\ee
where $(f_x,f_y,f_z) \approx  (\hbar vp_x/\Delta, \hbar v p_y/\Delta, m/\Delta)$, and we have re-written $\bar{\Phi}(p_x) = \int dx' e^{-i p_x x'} \Phi(x')$.
Recalling the identity
\be
\int_{-\infty}^\infty e^{iax} dx = e^{i\pi/4} \sqrt{\frac{\pi}{a}},\quad  a>0, 
\ee
where we have taken the principal value, and integrating out $p_x$ by completing the square in the argument of the exponential function, we obtain
\be
F_{0,z}(x,t) \approx \mathcal{N}  \sqrt{\frac{i \Delta}{ h v^2 t}} \Psi_t(x)  e^{i\Delta t/\hbar} %+ \mathcal{O}(t^{-3/2}), 
\label{eq:S-F}
\ee
where 
\be
%\Psi_t(x)=\int dx' \Phi(x') e^{-i \Delta (x-x')^2/(2t \hbar v^2)}
\Psi_t(x)=\int dx' \Phi(x+x') G_t(x'), \quad G_t(x) = e^{-i \Delta x^2/(2t \hbar v^2)}, 
\label{eq:S-Psit}
%\label{eq:Psit}
\ee
where we have changed dummy variables $x- x' \to x'$. We note that in obtaining the PD profile in both frozen and melting regimes, $F_{x,y}$ contributions are small (suppressed in the large $m\alpha/\hbar v \gg 1$ limit) and are negligible in $\la \rho \ra_{x,t}$. This is because while the integrand in $F_z$ is proportional to $m/\Delta \rightarrow 1$, the integrand in $F_x,F_y$ are proportional to $p_x,p_y$ respectively, which are much smaller than the gap size $m$. 

%and $\vec g = (1, 0, \hbar v p_y/\Delta, m/\Delta)$. The contributions $\mathcal{O}(t^{-3/2})$ \addJS{[JS: how do i see this order?]} arise from $f_x \approx \hbar vp_x/\Delta\red{\rightarrow i\hbar v/\Delta\,\partial_x}$ in Eq.~(\ref{eq:S-Fj})\red{, which creates additional powers of $t^{-1}$}. While they are the ``leading'' contributions to $F_{j=x} (x,t)$, as we will see, their contributions are small as compared to the other $F_{0,z}$ in the behavior of the probability density. . 

\subsection{Frozen Regime}

The imaginary Gaussian kernel in Eq.~(\ref{eq:S-Psit}) plays a crucial role in the dynamical evolution of the probability density. At short times, $G_t(x)$ exhibit spatial oscillations far faster than the spatial variations of $\Phi(x)$. Indeed, $\Phi(x)$ varies significantly for $x$ varying on order of the TBM width, $\alpha$. In contrast, $G_t(x)$ changes rapidly on length scales of order $\sqrt{2t\hbar v^2/ \Delta}$, and for short times, this length scale can be far shorter than $\alpha$. As a result, for short time windows (controllable by postquench $m$)
\be
0< t \ll \frac{\Delta \alpha^2}{2\hbar v^2} \approx \frac{m\alpha}{\hbar v} \tau, \quad \tau = \frac{\alpha}{2v}, 
\label{eq:timewindow}
\ee
we have $\Delta \alpha^2/(2t \hbar v^2) \gg 1$. 

In this short time window Eq.~(\ref{eq:timewindow}), $G_t\rightarrow e^{-i\pi/4}v\sqrt{2\pi\hbar t}\delta(\Delta x)$, and $\Phi(x+x')$ is effectively replaced by a constant $\Phi(x)$ times a factor containing $t^{1/2}$, and we can approximate
\be
\Psi_t(x) \approx \Phi(x) \int G_t(x') dx' = e^{-i\pi/4} \sqrt{\frac{2\pi \hbar v^2 t}{\Delta}}
\ee
yielding $F_{0,z} (x,t) =\mathcal{N} e^{i\Delta t/\hbar}$. As a result, at these short times $F_{0,z} (x,t)\approx \mathcal{N} \Phi(x) e^{i\Delta t/\hbar}$, so that $C(x,t) \approx \mathcal{N}\Phi(x) {\rm cos}(\Delta t/\hbar) $ and $S_z \approx \mathcal{N}\Phi(x) {\rm sin}(\Delta t/\hbar) $, where we have noted that $F_{x,y}$ are suppressed for large  $m \alpha/\hbar v$. 
%we have $\vec g \to (1, 0, 0, 1)$. 
Substituting into Eq.~(\ref{PD1m}) in the main text we obtain the Frozen profile 
\be \langle 
\rho\rangle_{t,x} \approx\mathcal{N}^2 |\Phi(x)|^2 
, \quad 0<t< \frac{\tau m \alpha}{\hbar v}.
\label{eq:S-frozen}
\ee
This frozen (i.e. near-stationary in time) profile closely conforms to that found from a direct numerical evaluation of the probability density found in Fig.2b of the main text. 

\subsection{Melting regime}

In the opposite regime to Eq.~(\ref{eq:timewindow}) (maintaining $m\alpha/\hbar v \gg 1$), $t \gg \frac{\Delta \alpha^2}{2\hbar v^2} = \frac{m\alpha}{\hbar v} \tau$, the imaginary Gaussian kernel in Eq.~(\ref{eq:S-Psit}) varies slowly in $x$. This will tend to spread $F_{0,z}(x,t)$ and hence the density of states out in space. The amplitude of $F_{0,z}(x,t)$ decays as $t^{-1/2}$ as in Eq.~(\ref{eq:S-F}). %Noting that large $m \alpha/\hbar v$, we have $\vec g \to (1, 0, 0, 1)$, 
We similarly obtain $C(x,t) \approx  [\Delta/(h v^2 t)]^{1/2} \mathcal{N} {\rm cos}(\Delta t/\hbar + \pi/4) \Psi_t(x)$  and $S_z \approx  [\Delta/(h v^2 t)]^{1/2} \mathcal{N} {\rm sin}(\Delta t/\hbar + \pi/4) \Psi_t(x)$. Note that $S_x,S_y\ll S_z$ because the values of $p_x,p_y$ that contribute to the integral giving $F_x,F_y$ are much smaller than $\Delta$ in the large $m\alpha/\hbar v \gg 1$ limit. 
%since large $p_x$ gives rise to a very rapidly oscillating integral and $p_y<\alpha^{-1}$ by definition. 
Plugging this into Eq.~(\ref{PD1m}) in the main text we obtain, the power-law melting profile found in Eq.~(\ref{eq:melting}) of the main text. 

Interestingly, in this regime the imaginary Gaussian kernel in Eq.~(\ref{eq:S-Psit}) varies slowly in $x$ far more slowly than $\Phi(x+x')$ varies in position. As a result, for $x \lesssim \alpha$ close to the domain wall $\Psi_t \approx \int \Phi(x+x') = \mathcal{N}^{-1}$. In this limit, we find that the probability density at different $x \lesssim \alpha$ collapse onto each other and decay as 
\be
\langle \rho\rangle_{t,x}  \to \frac{\Delta}{h v^2 t}, \quad t \gg \frac{m\alpha}{\hbar v} \tau, 
\label{eq:S-longtime}
\ee
displaying a universal $x$-independent power-law decay. Indeed, this collapse of PD at different $x$ positions in the long time limit is seen in a direct numerical evaluation in Fig. 2b of the main text. Indeed, we expect that at sufficiently long times, when $\Psi_t(x)$ has spread out, larger $x > \alpha$ within the width of $\Psi_t(x)$ will also similarly exhibit the universal $x$-independent power-law decay. 

\subsection{Illustration: exactly solvable example}

While we have obtained approximate expressions for $\Psi_t (x)$ [and hence, $F_j(x,t)$] above, it is instructive to illustrate the ``frozen'' and ``melting'' regime behavior through an exactly solvable model. To do so we consider the case where the initial TBM state $\Phi(x)$ takes the form of a Gaussian: $\Phi(x)=e^{-x^2/2\alpha^2}$, where $\alpha$ is the standard deviation. We can then obtain by direct integrating Eq.~(\ref{eq:S-Psit})%inverse variance. Then
\be%\begin{eqnarray}
\Psi_t(x) = \sqrt{\frac{hv^2 \alpha^2t}{\hbar v^2t + i\Delta \alpha^2}} \Big[{\rm exp} (-x^2/2\alpha^2)\Big]^{1+ \frac{\hbar v^2t}{\hbar v^2t + i\Delta \alpha^2}}
%\int e^{-i \Delta (x')^2/(2t \hbar v^2)} e^{-(x+x')^2/2\alpha^2}dx'\nonumber \\
%&=& \sqrt{\frac{\pi}{1-\frac{it \hbar v^2}{\alpha^2\Delta}}} e^{-\frac{x^2}{2\alpha^2}\frac{1}{1+\frac{it \hbar v^2}{\alpha^2\Delta}}}
%F(t) &\approx&  g_j\sqrt{\frac{\mathcal{N}\pi i \Delta}{t}}e^{i\Delta t}\int e^{-i\frac{\Delta (\Delta x)^2}{2t}}e^{-x'^2/2\alpha^2}dx'\notag\\
%&=& g_0\sqrt{\frac{2\pi\mathcal{N} }{1-\frac{2itb}{\Delta}}}e^{i\Delta t}e^{-\frac{bx^2}{1+\frac{2itb}{\Delta}}}
\label{example}
\ee %\end{eqnarray}
Evidently, when $t \ll \frac{\alpha^2 \Delta}{\hbar v^2}$ then we have
\be
\Psi_t(x) \to \sqrt{\frac{hv^2t}{i\Delta}} \Big[{\rm exp} (-x^2/2\alpha^2)\Big], \quad t \ll \frac{\alpha^2 \Delta}{\hbar v^2}
\ee
Substituting in Eq.~(\ref{eq:S-F}), we obtain $F_j(x,t) \approx g_j e^{i\Delta t} \mathcal{N} {\rm exp} (-x^2/2\alpha^2) \propto \Phi(x) e^{i\Delta t} $ that leads to the frozen (near-stationary in time) PD profile as discussed above. 

In the opposite limit $t \gg \frac{\alpha^2 \Delta}{\hbar v^2}$, 
\be
\Psi_t(x) \to \sqrt{2\pi \alpha^2} \Big[{\rm exp} (-x^2/\alpha^2)\Big], \quad t \gg \frac{\alpha^2 \Delta}{\hbar v^2}
\ee
Recalling that $\sqrt{2\pi \alpha^2} = \mathcal{N}^{-1}$ is the inverse normalization constant for a Gaussian profile $\Phi(x)=e^{-x^2/2\alpha^2}$, and substituting into Eq.~(\ref{eq:S-F}), we similarly obtain the power-law decay in the PD described in Eq.~(\ref{eq:S-longtime}) (in the limit $x \ll \alpha$). Note that the Gaussian profile above is special in that it does not spread due to its covariance under convolution by a Gaussian Kernel. For more general profiles, we will expect spread effects like what shown in Fig. 1 of the main text, as well as that shown in 
Figs.~\ref{plots} and \ref{plots2} of the Supplementary Information.

\section{Precession and dynamical equations of motion}

Although we have already provided the full solution to the PD and pseudospin expectation, it is still instructive to understand how the dynamical equations governing these quantities are related to the well-known spin precession equation. Due to the lack of translation invariance in the initial state, the evolution is not diagonal in momentum space, and the equations of motion will involve projectors.
%We next compute the time derivative of the LDOS $\langle \rho\rangle_{t,\bold r}=\langle \psi_{p_y}(t,\bold r)| \psi_{p_y}(\bold r, t)\rangle$ and pseudospin expectation $\langle \boldsymbol\sigma\rangle_{t,\bold r}=\langle \psi_{p_y}(t,\bold r)|\boldsymbol\sigma| \psi_{p_y}(\bold r, t)\rangle$ in more explicit detail than in the main text, such as to reveal the origin of momentum-space non-locality. 
With $|\psi_{p_y}(t,\bold r)\rangle = |\bold r\rangle\langle \bold r|e^{-i\hat{\vec H}_ft}|\Psi^{(0)}\rangle$, the rate of change of the PD can be directly evaluated from Eq.~(\ref{PD0}) of the main text and takes the form
\begin{eqnarray}
\frac{d\langle \rho\rangle_{t,\bold r}}{dt} %&=& \frac{d}{dt}\langle\psi_{p_y}(t,\bold r)|\psi_{p_y}(t,\bold r)\rangle\notag\\
%&=&\frac{d}{dt}\left|\langle \bold r|e^{-i\hat{\vec H}_ft}|\Psi^{(0)}\rangle\right|^2\notag\\
%&=& \frac{d}{dt}\langle \Psi^{(0)}|e^{i\hat{\vec H}_ft}P_{\bold r}\,e^{-i\hat{\vec H}_ft}|\Psi^{(0)}\rangle\notag\\
&=& i\langle \Psi^{(0)}|e^{i\hat{\vec H}_ft}\left[\hat{\vec H}_f,\hat{\vec{P}}_{\bold r}\right]e^{-i\hat{\vec H}_ft}|\Psi^{(0)}\rangle\notag\\
%&=& 2\,\text{Im}\langle \Psi^{(0)}|e^{i\hat{\vec H}_ft}P_{\bold r}\hat{\vec H}_f\,e^{-i\hat{\vec H}_ft}|\Psi^{(0)}\rangle\notag\\
%&=&2\,\text{Im}\langle P_{\bold r}\,\hat{\vec H}_f\rangle_{t}\notag\\
&=& 2\,\text{Im}\langle (\hat{\vec{P}}_\bold r\,\tilde{\bold d})\cdot\boldsymbol\sigma\rangle_{t},
%&=&\sum_{p_x,p'_x}e^{ix(p_x-p_x')}\frac{d}{dt}\langle \Psi^{(0)}(\bold r)|e^{i\hat{\vec H}_f(\bold p')t} e^{-i\hat{\vec H}_f(\bold p)t}|\Psi^{(0)}(\bold r)\rangle \notag\\
%&=& i\frac{\mathcal{N}^2}{2}\sum_{p_x,p'_x}e^{ix(p_x-p_x')}\bar\Phi^*( p'_x)\bar\Phi(p_x)\langle\phi_0|e^{i\hat{\vec H}_f(\bold p')t}[\hat{\vec H}_f(\bold p')- \hat{\vec H}_f(\bold p)] e^{-i\hat{\vec H}_f(\bold p)t}|\hat\phi_0\rangle.
%&=& \frac{\mathcal{N}^2}{2}\sum_{p_x,p'_x}e^{ix(p_x-p_x')}\bar\Phi^*(p'_x)\bar\Phi(p_x)\langle\phi_0|e^{i\hat{\vec H}_f(\bold p')t}[i(\bold d(\bold p')-\bold d(\bold p))\cdot \sigma] e^{-i\hat{\vec H}_f(\bold p)t}|\hat\phi_0\rangle \notag\\
%&=&\frac{\mathcal{N}^2}{4}\sum_{p_x,p'_x}\left[e^{ix(p_x-p_x')}\bar \Phi(p')^*\bar\Phi(p)+e^{-ix(p_x-p_x')}\bar \Phi(p)^*\bar\Phi(p')\right]\bold m(p,p',t)
%&=&2\,\text{Im} \langle \psi_{p_y}(t,\bold r)|\sigma\cdot\tilde{\bold d}|\psi_{p_y}(t,\bold r)\rangle\notag\\
%+2\,\text{Im} \langle \psi_{p_y}(t,\bold r)|\tilde{\bold d}|\psi_{p_y}(t,\bold r)\rangle
\label{rhoxtm2}
\end{eqnarray}
where $\hat{\vec{P}}_{\bold r}=|\bold r\rangle\langle \bold r|$ is projection operator onto the real-space state $|\bold r\rangle$ at position $\bold r$, $\tilde{\bold d}=\sum_\bold p \bold d(\bold p )|\bold p \rangle \langle \bold p |$ is the projection operator onto the spinor eigenstate corresponding to $\bold d(\bold p )$ and $\langle ... \rangle_t$ denotes an expectation evaluated with the evolved initial state at time $t$ (after the quench). 

The projection $\hat{\vec{P}}$ enters because the expectation was evaluated at the particular position $\bold r$, not across the entire wavefunction $\psi_{p_y}(t,\bold r)$. Importantly, the simultaneous presence of operators $\hat{\vec{P}}$ and $\hat{\vec H}_f$, which are respectively diagonal in real-space and momentum-space, leads to a nontrivial imaginary part of the expectation $\text{Im}\langle P_{\bold r}\,\hat{\vec H}_f\rangle_{t}$, even though $\text{Im}\langle \hat{\vec H}_f\rangle_{t}=0$ due to the Hermiticity of $\hat{\vec H}_f$. Indeed the rate of change $\frac{d\langle \rho\rangle_{t,\bold r}}{dt}$ at the specific position $\vec r$ can be construed as a measure of the ``non-Hermiticity'' of $P_\bold r \hat{\vec H}_f$, which is manifested through the imaginary part of its eigenvalues. This exhibits how the probability density at position $\vec r$ changes over time evolving to other positions.

% signature of the 

Analogously, the rate of change of the pseudospin expectation is given by
\begin{eqnarray}
\frac{d\langle \boldsymbol\sigma\rangle_{t,\bold r}}{dt} 
&=& i\langle \Psi^{(0)}|e^{i\hat{\vec H}_ft}\left[\hat{\vec H}_f,\hat{\vec{P}}_{\bold r} \boldsymbol\sigma P_\bold r\right]e^{-i\hat{\vec H}_ft}|\Psi^{(0)}\rangle\notag\\
&=& 2\,\text{Re}\,\langle (\hat{\vec{P}}_{\vec r}\,\tilde{\bold d})\times\boldsymbol\sigma\rangle_{t}.
\label{sigmaxtm2}
\end{eqnarray}
This is reminiscent of the usual precession equation with RHS given by $2\,\tilde{\bold d}\times\boldsymbol\sigma$, except that the $\hat{\vec{P}}_{\vec r}$ projector now appear prominently. Comparing Eqs.~\ref{rhoxtm2} and \ref{sigmaxtm2}, we see that the PD and pseudospin evolutions are both governed by $\hat{\vec{P}}_\bold r \tilde{\bold d}$. Explicitly, the effect of $\hat{\vec{P}}_\bold r$ is to render this operator product non-local in momentum space:
\be
\hat{\vec{P}}_\bold r\,\tilde{\bold d} =\sum_\bold p\bold d(\bold p)| \bold r\rangle\langle \bold r|\bold p\rangle\langle \bold p| %\notag\\
= \sum_{\bold p,\bold p'}\bold d(\bold p)| e^{i(\bold p'-\bold p)\cdot \bold r}|\bold p'\rangle\langle \bold p|.
\ee
Due to the spatial inhomogeneity of the real-space profile of the initial state, both $\bold p$ and $\bold p'$ are not connected by a delta function, resulting in a more complicated expression involving both $\bold p$ and $\bold p'$. This can be interpreted as an interference mechanism between different momentum components, which gives rise to time dependence in the PD as well as the pseudospin expectation. As we discuss in the main text, this interference leads to a ``Wick-rotated'' diffusion process, by which slow decay emerges in the melting regime of both $\la \rho\ra_{x,t}$ and $\la \boldsymbol{\sigma}\ra_{x,t}$ arise. 
 
\begin{comment}
To be more concrete, we expand the final expressions in Eqs.~\ref{rhoxtm2} and \ref{sigmaxtm2} in terms of contributions from the initial TBM profile $\bar\Phi(p_x)$:
\begin{eqnarray}
\frac{d\langle \rho\rangle_{t,\bold r}}{dt} &=&  i\frac{\mathcal{N}^2}{2}\sum_{p_x,p'_x}e^{ix(p_x-p_x')}\bar\Phi^*( p'_x)\bar\Phi(p_x)\langle\phi_0|e^{i\hat{\vec H}_f(\bold p')t}[\hat{\vec H}_f(\bold p')- \hat{\vec H}_f(\bold p)] e^{-i\hat{\vec H}_f(\bold p)t}|\hat\phi_0\rangle,
\end{eqnarray}
\begin{eqnarray}
\frac{d\langle \boldsymbol\sigma\rangle_{t,\bold r}}{dt}&=& \frac{\mathcal{N}^2}{2}\sum_{p_x,p'_x}e^{ix(p_x-p_x')}\bar\Phi^*(p'_x)\bar\Phi(p_x)\langle\phi_0|e^{i\hat{\vec H}_f(\bold p')t}[(\bold d(\bold p)+\bold d(\bold p'))\times \boldsymbol\sigma ] e^{-i\hat{\vec H}_f(\bold p)t}|\hat\phi_0\rangle.
\label{rhoprofile}
\end{eqnarray}
\end{comment}

\subsection{Relation to Bloch dynamics for single spins}
One can relate the above results to the much simpler well-known precession equation
\begin{equation}
\frac{d \tilde{\vec{S}}(\bold p,t)}{dt} = 2\, \tilde{\bold{S}}(\bold p,t)\times \bold d(\bold p),
\label{sigmaxtm3m}
\end{equation}
by defining a magnetization i.e. spectral pseudospin denstiy function $\tilde{\bold S}(\bold p,t)$  via $\langle \boldsymbol \sigma (t)\rangle = \pi \mathcal{N}^2 \sum_{p_x}|\bar\Phi(p_x)|^2\tilde{\bold S}(\bold p,t)$, where $\langle \boldsymbol\sigma(t)\rangle = \int \langle \boldsymbol\sigma\rangle_{t,x}\,dx$ is the total pseudospin expectation for the entire system. With the spatial inhomogeneity integrated over, we indeed obtain an equation of motion that is local in momentum.

\clearpage
\newpage

\renewcommand{\theequation}{S-\arabic{equation}}
\renewcommand{\thefigure}{S-\arabic{figure}}
\renewcommand{\thetable}{S-\Roman{table}}
\makeatletter
\renewcommand\@biblabel[1]{S#1.}
\setcounter{equation}{0}
\setcounter{figure}{0}

\twocolumngrid

\section{Supplementary Information for
``Quenched topological boundary modes can persist in a trivial system''}

\section{Illustration: Dynamical Evolution of Pseudospin Expectation Density}
\begin{figure*}
\centering
\includegraphics[scale=.29]{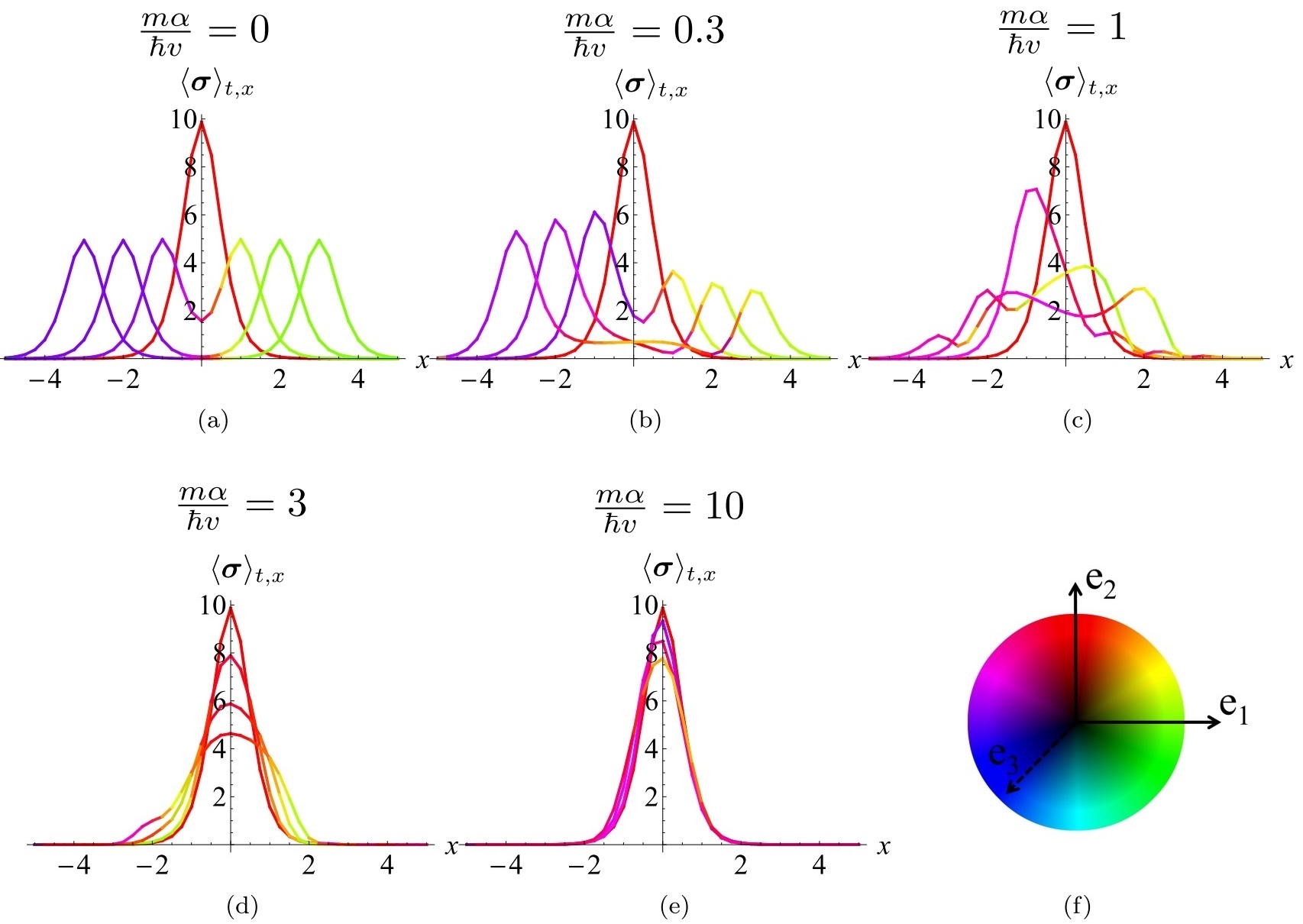}
\caption{Time evolution of $\langle \boldsymbol\sigma \rangle_{t,x}$ for $m\alpha/\hbar v\neq 0$, $p_y=0$ from an initial $x=0$ spike, with profiles at $t=1,2,3$ being increasingly spread out or attentuated. The pseudospin direction is color coded according to the color wheel in (f). For the $p_y=0$ case at hand, the evolution is restricted to the $\hat{\bold e}_1$-$\hat{\bold e}_2$ plane. While it initially spreads out unattenuated for small $m\alpha/\hbar v$, for larger $m\alpha/\hbar v$ attenuation occurs and the spins takes longer to relax to $\pm\hat{\bold e}_1$. Finally for large $m\alpha/\hbar v\gg 1$, the system exhibits frozen followed by melting behavior, with diminishing group velocity and attenuation. Positions are given in units of lattice spacing $a$, and times $t$ in units of $v/\alpha$.   
}
\label{plots}
\end{figure*}

\begin{figure*}
\centering
\includegraphics[scale=.29]{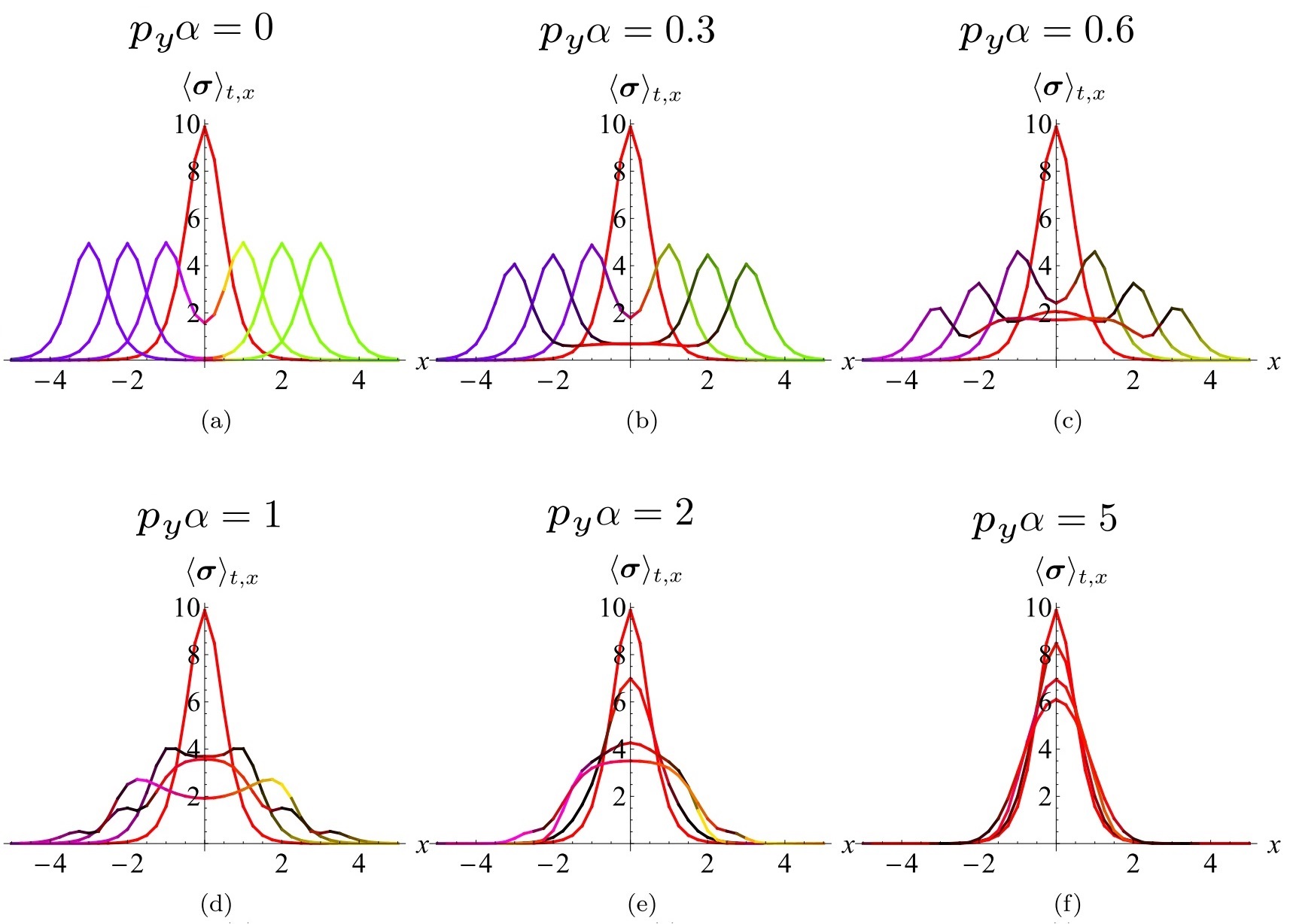}
\caption{Time evolution of $\langle \boldsymbol\sigma (x,t)\rangle$ for $p_y\alpha\neq 0$, $m\alpha/\hbar v=0$ from an initial $x=0$ spike, with profiles at $t=1,2,3$ being increasingly spread out or attentuated. The main differences with the $m\alpha/\hbar v\neq 0,p_y=0$ case (Fig. \ref{plots} above) is that the attenuation occurs symmetrically and the spin evolution cants out of the plane with nonzero $\hat{\bold e}_3$ component, indicated by darker shades (see color legend in Fig. \ref{plots}(f)). At large $p_y\alpha$, the system also exhibits frozen followed by melting behavior. Positions are given in units of lattice spacing $a$, and times $t$ in units of $v/\alpha$.    
}
\label{plots2}
\end{figure*}

In this section we show the dynamical evolution of the full pseudospin expectation density which, as discussed in the main text, involves a complex intertwining of precession and interference between wave components composing the spatially localized profile of the TBM state. Concentrating on quenches into a Dirac-type Hamiltonian as discussed in the main text, we plot the dynamical evolution in Figs.~\ref{plots} and \ref{plots2} for special but important cases $p_y=0,m\neq 0$ and $m=0,p_y\neq 0$ respectively. 

When $m=p_y=0$ (exactly solved in the final Supplementary section), a state prepared as a TBM prequench with $\la \boldsymbol{\sigma}\ra \propto \hat{\bold{e}}_2$ starts at $x=0$ splits into two oppositely traveling wavepackets with $\la \boldsymbol{\sigma}\ra \propto \pm\hat{\bold{e}}_1$. This light-cone like spreading without attenuation is the result of the absence of any time scale/energy scale imposed by nonzero $m$ or $p_y$. 

When $p_y=0$, the pseudospin expectation never acquires any $\hat{\bold e}_3$ component. As the postquench gap $m$ increases, attenuation occurs and the spins takes longer to relax to $\pm\hat{\bold e}_1$, although the integrated PD over all space of course remains conserved at unity. As shown in Fig.~\ref{plots}(b) and (c), the attenuation is asymmetric, and the initial decay on one side can be even slower than the case of $m=0$. However, when $m\alpha/\hbar v$ increases above unity, the decay becomes rapid while the group velocity and spin relaxation becomes very slow. Finally, at large $m\alpha/\hbar v\gg 1$, the wavepacket becomes very spatially localized and behaves almost like a single Bloch spinor with negligible spatial inhomogeneity [Fig.~\ref{plots}(e)]. This is also the regime where it exhibits frozen and then melting behavior about the initial domain wall $x=0$, as discussed extensively in the main text, as well as in the following section. This is akin to an imaginary damped precessing pseudospin.

When $m=0$ instead, the attenuation occurs symmetrically and the evolved pseudospin cants out of the plane with nonzero $\hat{\bold e}_3$ component [Fig.~\ref{plots2}]. This occurs because the precession axis no longer all lie within the same plane across different momentum sectors. 

Interestingly, when $p_y\alpha$ increases above unity [Fig.~\ref{plots2}], the decay becomes very rapid like in the $m\neq 0$ case. At large $p_y\alpha$, the pseudospin evolution proceeds similarly as in the large $m,p_y=0$ case, exhibiting a frozen and then melting behavior. This is because such behavior results fundamentally from large $\alpha\Delta/\hbar v=\alpha\sqrt{m^2/\hbar^2v^2+p_y^2}$, which can result from either large $m\alpha/\hbar v $ or large $p_y \alpha$ (or both). 

\newpage

\section{Pseudospin Expectation Density in the melting regime for arbitrary $p_y$} 

In the following, we analyze the pseudospin expectation density in the melting regime when $m\alpha/\hbar v\gg 1$. While in earlier sections we chose $p_y$ small, in this section we relax the assumption that $|p_y|<\alpha^{-1}$, in order to extend our scope to arbitrary electronic fillings. As we shall see, power-law behavior shall still emerge asymptotically even when $p_y$ is comparable to $\Delta$. We shall temporarily work in units of $\alpha,\hbar, v=1$ for notational brevity.\\

\subsection{Pseudospin Expectation $\la \boldsymbol \sigma\ra_{t,x}$}

We now explicitly derive the pseudospin expectation behavior in the regime $\Delta=\sqrt{m^2+p_y^2}\gg \alpha^{-1}$, specifically to leading order in $\Delta^{-1}$. 
Firstly, note that $\hat{\vec d}(p_x)$ can be approximated as
\begin{equation}
\hat{\vec d}(p_x)=\frac1{\sqrt{\Delta^2+p_x^2}}\left(\begin{matrix}
  p_x \\
  p_y\\
  m
  \end{matrix}\right)\approx\frac1{\Delta}\left(\begin{matrix}
  p_x \\
  p_y\\
  m,
  \end{matrix}\right)
	\rightarrow\frac1{\Delta}\left(\begin{matrix}
  -i\partial_x \\
  p_y\\
  m,
  \end{matrix}\right)
\end{equation}
where $ip_x$ has been replaced by a derivative operator on the profile $\Phi(x)$. For simplicity of notation in the following derivations, we shall introduce 
\begin{eqnarray}
\Theta_t(x)&=&\sqrt{\frac{\Delta}{\hbar v^2t}}\Psi_t(x)\notag\\
&=&\sqrt{\frac{\Delta}{\hbar v^2t}}\int dx'\Phi(x+x')e^{-i\Delta x^2/(2t\hbar v^2)}\notag\\
\end{eqnarray}
and the corresponding convolution with $\Phi'(x)$:
\begin{eqnarray}
\Theta'_t(x)&=&\sqrt{\frac{\Delta}{\hbar v^2t}}\int dx'\Phi'(x+x')e^{-i\Delta x^2/(2t\hbar v^2)}.\notag\\
\end{eqnarray}
This gives us
\begin{subequations}
\begin{equation}
C\approx\frac{\mathcal{N}}{\sqrt{2}}\text{Re}[\Theta_t(x)e^{i\Delta t}]=\frac{\mathcal{N}}{\sqrt{2}}|\Theta_t(x)|\cos\gamma(x,t)\,
\end{equation}
\begin{eqnarray}\bold S&\approx& \frac{\mathcal{N}}{\sqrt{2}\Delta}\left[ \left(\begin{matrix}
 0 \\
  p_y\\
  m
  \end{matrix}\right)\text{Im}[\Theta_t(x)e^{i\Delta t}]-\left(\begin{matrix}
 1 \\
  0\\
  0
  \end{matrix}\right)\text{Im}[i\Theta'_t(x)e^{i\Delta t}] \right]\notag\\
	&=&\frac{\mathcal{N}}{\sqrt{2}\Delta}\left(\begin{matrix}
 -|\Theta'_t(x)|\cos\gamma'(x,t) \\
  p_y \,|\Theta_t(x)|\sin\gamma(x,t)\\
  m\,|\Theta_t(x)|\sin\gamma(x,t)
  \end{matrix}\right)
	\end{eqnarray}
  \label{CSd}
  \end{subequations} 
\noindent where we have defined $\gamma(x,t)=\Delta t + \eta(x)$ and $\gamma'(x,t)=\Delta t + \eta'(x)$, and $\Theta_t(x)=|\Theta_t(x)|e^{i\eta(x)}$ and $\Theta'_t(x)=|\Theta'_t(x)|e^{i\eta'(x)}$, i.e. $\tan \eta (x)=\text{Im}[\Theta_t(x)]/\text{Re}[\Theta_t(x)]$. Then upon subtitution in Eq. \ref{sigmaxt}, and noting that $\text{Im}[\bold S]=0$, we get 
\begin{widetext}
\begin{eqnarray}
\frac{2}{\mathcal{N}^2}\langle \boldsymbol\sigma\rangle_{t,x}^\text{}&=&\frac{2}{\mathcal{N}^2}\left[\hat{\vec{s}} (|C|^2-|\bold S|^2) +2\text{Re}[\bold S^*C]\times \hat{\vec{s}}+2\text{Re}[(\bold S^*\cdot\hat{\vec{s}})\bold S]\right]\notag\\
&=& \hat{\bold e}_2\left(|\Theta_t(x)|^2  \cos 2 \gamma(x,t)-\frac1{\Delta^2} |\Theta'_t(x)|^2  \cos^2 \gamma'(x,t)\right)
+\frac{2}{\Delta}\left(\begin{matrix}
 -|\Theta'_t(x)|\cos\gamma'(x,t) \\
  p_y \,|\Theta_t(x)|\sin\gamma(x,t)\\
  m\,|\Theta_t(x)|\sin\gamma(x,t)
  \end{matrix}\right)|\Theta_t(x)|\cos\gamma(x,t)\times \hat{\bold e}_2\notag\\
	&&+\frac{2}{\Delta^2}p_y|\Theta_t(x)|\sin\gamma(x,t)\left(\begin{matrix}
 -|\Theta'_t(x)|\cos\gamma'(x,t) \\
  p_y \,|\Theta_t(x)|\sin\gamma(x,t)\\
  m\,|\Theta_t(x)|\sin\gamma(x,t)
  \end{matrix}\right)\notag\\
	\begin{comment}
	&=&\left(\begin{matrix}
 0 \\
 |\Theta_t(x)|^2  \cos 2 \gamma(x,t)-\frac1{\Delta^2} |\Theta'_t(x)|^2  \cos^2 \gamma'(x,t)\\
  0
  \end{matrix}\right)+\frac{2}{\Delta}|\Theta_t(x)|\cos\gamma(x,t)\left(\begin{matrix}
 -m\,|\Theta_t(x)|\sin\gamma(x,t)\\
  0\\
  -|\Theta'_t(x)|\cos\gamma'(x,t) 
  \end{matrix}\right)\notag\\
	&&+\frac{2}{\Delta^2}p_y|\Theta_t(x)|\sin\gamma(x,t)\left(\begin{matrix}
 -|\Theta'_t(x)|\cos\gamma'(x,t) \\
  p_y \,|\Theta_t(x)|\sin\gamma(x,t)\\
  m\,|\Theta_t(x)|\sin\gamma(x,t)
  \end{matrix}\right)\notag\\
\end{comment}
&=& \left[|\Theta_t(x)|^2\left(1-\frac{2m^2\sin^2\gamma(x,t)}{\Delta^2}\right)-|\Theta'_t(x)|^2\frac{\cos^2\gamma'(x,t)}{\Delta^2}\right]\hat{\bold e}_2\notag\\
&&-\frac{2}{\Delta}|\Theta_t(x)|  \left(\begin{matrix}
 m\cos \gamma(x,t)\sin\gamma(x,t)|\Theta_t(x)|+\frac{p_y}{\Delta}\sin\gamma(x,t)\cos\gamma'(x,t)|\Theta'_t(x)| \\
  0 \\
  -\frac{p_y}{\Delta}m\sin^2\gamma(x,t)|\Theta_t(x)|+\cos \gamma(x,t)\cos\gamma'(x,t)|\Theta'_t(x)|
  \end{matrix}\right )
	\label{large_mky}
\end{eqnarray}
\end{widetext}
where $\hat{\vec{s}} = \hat{\bold e}_2$ is the original pseudospin polarization of the state prepared in the TBM. In final expression, we see that the component in the direction of $\hat{\vec{s}}$ originates not just from the convolution $\Theta_t(x)$ involving the initial state profile, but also $\Theta'_t(x)$ which involves the convolution with the \emph{spatial derivative} of the initial state profile. This is the full solution of the dynamical evolution in the $\Delta\gg \alpha^{-1}$ limit, with the first term of the final line indicating the extent of pseudospin expectation persistence. 

To make further headway, we shall define
\begin{equation}
m_t(x)=-\frac{\Theta'_t(x)}{\Theta_t(x)}
\label{mtx}
\end{equation}
which contains information about the nontrivialty of the imaginary Gaussian kernel. Firstly, in the frozen regime of small $t$, the kernel tends to a delta function, and $\Theta'_t(x)$ becomes genuinely the derivative of $\Theta_t(x)$. Hence for small $t$, $m_t(x)|_{t\approx 0} \rightarrow -\frac{d}{dx}\log e^{-\int^xM(x')dx'}=M(x)$, the initial mass gap distribution given by the Jackiw-Rebbi solution. Hence $m_t(x)$ can be regarded as a ``time-relaxed'' mass gap distribution which equilibrates to zero after some time. But for large $t$, the definition Eq.~\ref{mtx} permits no further simplification due to imaginary Gaussian Kernel. Nevertheless, the large $t$ limit can still be reduced as follows:
\begin{widetext}
\begin{eqnarray}
2\pi\langle \boldsymbol\sigma\rangle_{t\rightarrow\infty,x}^\text{}&\rightarrow & \frac{\pi\Delta \mathcal{N}^2}{t}|\Psi_t(x)|^2\left[1-\frac{2m^2\sin^2\gamma(x,t)+m_t(x)^2\cos^2\gamma(x,t)}{\Delta^2}\right]\hat{\bold e}_2\notag\\
&&-\frac{2\pi \mathcal{N}^2}{t}|\Psi_t(x)|^2 \left(\begin{matrix}
 \cos \gamma(x,t)\sin\gamma(x,t)\left(m-m_t(x)\frac{p_y}{\Delta}\right) \\
  0 \\
  -\frac{p_y}{\Delta}m\sin^2\gamma(x,t)-m_t(x)\cos^2 \gamma(x,t)
  \end{matrix}\right )\notag\\
	&=&-\frac{\pi \mathcal{N}^2}{t}|\Psi_t(x)|^2 \left(\begin{matrix}
 2\cos \gamma(x,t)\sin\gamma(x,t)\left(m-m_t(x)\frac{p_y}{\Delta}\right) \\
  \frac1{\Delta}\left(2m^2\sin^2\gamma(x,t)+m_t(x)^2\cos^2\gamma(x,t)\right)-\Delta \\
  -\frac{2p_y}{\Delta}m\sin^2\gamma(x,t)-2m_t(x)\cos^2 \gamma(x,t)
  \end{matrix}\right )
\label{large_mky2}
\end{eqnarray}
\end{widetext}
If we recall the assumption $|p_y|<\alpha^{-1}$ i.e. that the occupied states are in the pre-quench gap, $\Delta\gg \alpha^{-1}$ also implies that $p_y\ll \Delta$, $m\approx \Delta$ and $m_t(x)<\alpha^{-1}\gg \Delta$. Hence Eq.~\ref{large_mky2} further simplifies to 
\begin{eqnarray}
\langle \boldsymbol\sigma\rangle_{t\rightarrow \infty,x}^\text{}&\approx & \frac{\Delta \mathcal{N}^2}{2t}|\Psi_t(x)|^2\cos 2\gamma(x,t)\,\hat{\bold e}_2\notag\\
&&-\frac{ \mathcal{N}^2}{2t}|\Psi_t(x)|^2 
m\sin 2\gamma(x,t) \hat{\bold e}_1\notag\\
	&\approx&-\frac{ m\mathcal{N}^2}{2t}|\Psi_t(x)|^2 \left(\begin{matrix}
 \cos 2mt \\
  \sin 2mt \\
  0
  \end{matrix}\right )
\label{large_mky3},
\end{eqnarray}
which is the precession behavior in the melting regime as discussed in the main text.

%Eq.~\ref{large_mky3} reveals a simple precession behavior in the asymptotic limit of large $m$ and $t$. Notably, the precession occurs solely in the $\hat{\bold e}_1$-$\hat{\bold e}_2$ plane with recurrence interval $\Delta t= \pi/m$, even though  $\langle \boldsymbol\sigma\rangle$ for the bulk eigenstates after the quench have no component along $\hat{\bold e}_2$. Another key aspect of its behavior is its $\frac1{t}$ power-law decay in the large $t$ limit, which fundamentally arises from the Gaussian kernel $g(x)$ (Eq.~\ref{kernel}) resulting from the large $\Delta$ expansion of the energy $\epsilon_p=\sqrt{\Delta^2+p_x^2}\approx \Delta+\frac{p_x^2}{2\Delta}$. This behavior hinges on the quadratic ($p_x^2$) form of the large $\Delta$ expansion; alternative forms will have resulted in very different, if not non-analytic or non-convergent, behaviors. 
% half of that of nonzero $p_y$ because the latter case involves a non-planar precession trajectory that closes only every period of $2\pi/m$.

To summarize the above derivations, we first identified the large $\Delta$ condition for slow power-law decay of the key quantities $C$ and $\bold S$. Next, the full expression for $\langle\boldsymbol\sigma\rangle^\text{}_{t,x}$ is derived in this limit (Eq.~\ref{large_mky}). The large $t$ limit is then taken, resulting in a simplification to Eq.~\ref{large_mky2}. Finally, the assumption $|p_y|<\alpha^{-1}$ on the initial state occupancy is invoked to result in the even simpler expression Eq.~\ref{large_mky3}, as borne out in the melting regime of Fig.~3 of the main text. 
\\

\section{Explicit analytic example: exact results for the $m=p_y=0$ case}
\label{exp1}

In this final section, we provide pedagogical examples of the analytically tractable limit of $m=p_y=0$. It was shown in Fig.~\ref{plots}(a) that in this special limit, the initial wavepacket does not decay at all, but instead ``splits'' into two and depart in equal and opposite velocities. Here we shall show how this behavior can be derived.

The post-quench Hamiltonian $p_x\sigma_x$ is gapless with eigenenergy $\varepsilon_{p_x}=|p_x|$, and all its bulk states are energetically accessible from any initial TBM state. Its exact linear dispersion also admits an exact analytic solution for the special choice of initial TBM state given by our ansatz $\bar\Phi(p_x)=\pi\alpha\,\text{sech}\frac{\pi\alpha p_x}{2}$. The post-quench state evolution can be computed from Eq.~S-1 (still working with units $\hbar v =1$):
\begin{widetext}
\begin{eqnarray}
\psi^{m=0}_{p_y=0}(x,t)
&=& \mathcal{N}\sum_{p_x,p_y,\pm} \delta_{p_y,0}e^{i\bold p \cdot \bold x}e^{\mp i\epsilon_{\bold p}t}\bar\Phi(p_x)\frac{\mathbb{I}\pm \hat{\bold d}(\bold p)\cdot \boldsymbol\sigma}{2}\hat{\bold s}\notag\\
&=& \frac{\mathcal{N}}{2\sqrt{2}}\sum_{p_x,\pm} e^{ip_xx}e^{\mp i|p_x|t}\bar\Phi(p_x)\left[\left(\begin{matrix}1 \\ i\\
\end{matrix}\right)\pm \text{ sgn}(p_x)\left(\begin{matrix} i\\  1\\
\end{matrix}\right)\right]\notag\\
&=& \frac{\mathcal{N}\pi\alpha}{2}\sum_{p_x} \frac{e^{ip_xx}}{\cosh\frac{\pi \alpha p_x}{2}}\left[\cos (|p_x|t)\left(\begin{matrix}
  1\\
  i\\
\end{matrix}\right)+\text{sgn}(p_x)\sin(|p_x|t)\left(\begin{matrix}
  1\\
  -i\\
\end{matrix}\right)\right].
\end{eqnarray}
On the first line, we have projected the initial spinor into the upper and lower bands, where time evolution is given by phase rotations. This is then simplified in terms of a sum of $p_x$, which will be converted into an integral below:
\begin{eqnarray}
\psi^{m=0}_{p_y=0}(x,t)&=& \sqrt{2}\mathcal{N}\int_{-\infty}^\infty  \frac{e^{iPX}}{\cosh P}\left[\cos (|P|T)\left(\begin{matrix}
  1\\
  i\\
\end{matrix}\right)+\sin (PT)\left(\begin{matrix}
  1\\
  -i\\
\end{matrix}\right)\right]dP\notag\\
&=& \frac{\pi \mathcal{N} }{\sqrt{2}}\left[\left(\text{sech}\frac{\pi(T-X)}{2}+\text{sech}\frac{\pi(T+X)}{2}\right)\left(\begin{matrix}
  1\\
  i\\
\end{matrix}\right)+i\left(\text{sech}\frac{\pi(T-X)}{2}-\text{sech}\frac{\pi(T+X)}{2}\right)\left(\begin{matrix}
  1\\
  -i\\
\end{matrix}\right)\right]\notag\\
&=&\frac{1}{2\sqrt{\alpha}\cosh\frac{t+x}{\alpha}\cosh\frac{t-x}{\alpha}} \left(\begin{matrix}
	\cosh\frac{x}{\alpha}\cosh\frac{t}{\alpha}+i\sinh\frac{x}{\alpha}\sinh\frac{x}{\alpha}	\\
	i\cosh\frac{x}{\alpha}\cosh\frac{t}{\alpha}+\sinh\frac{x}{\alpha}\sinh\frac{x}{\alpha}	\\		
\end{matrix}\right)
\label{evo2}
\end{eqnarray}
\end{widetext}
where $X=\frac{2}{\pi\alpha}x$, $T=\frac{2}{\pi\alpha}t$ are the spatial and temporal displacements in units of $\frac{\pi}{2}\alpha$. The integral is, with this fortuitous choice of profile $\bar\Phi(p_x)$, analytically tractable with contour integration. 

At long $t$, we observe manifest exponential decay behavior $\psi^{m=0}_{p_y=0}(x,t)\sim e^{-t/\alpha}$ with a timescale of $\alpha$. 
This scale arises from the spatial decay of the initial mass profile $m_t(x)$, which should govern the post-quench dynamics because there is no other scale introduced. In the massive ($m\neq 0$) post-quench case, $m$ will also determine the post-quench dynamics. Unlike in the special large $m$ case, the initial state generically decays exponentially according to timescales set by the energy scales involved.   

One can directly compute the pseudospin expectation of this $m=p_y=0$ case:
\begin{subequations}
\begin{eqnarray}\langle\sigma_x\rangle_{t,x}&=&\frac1{4\alpha}\left(\text{sech}^2\frac{t-x}{\alpha}-\text{sech}^2\frac{t+x}{\alpha}\right)\notag\\
&=&\frac{\sinh\frac{2x}{\alpha}\sinh\frac{2t}{\alpha}}{\alpha\left(\cosh\frac{2t}{\alpha}+\cosh\frac{2x}{\alpha}\right)^2}\end{eqnarray}
\begin{equation}\langle \sigma_y\rangle_{t,x}=\frac1{2\alpha\cosh\frac{t-x}{\alpha}\cosh\frac{t+x}{\alpha}}\end{equation}
\begin{equation}\langle \sigma_z\rangle_{t,x}=\langle \psi(x,t)|\sigma_3|\psi(x,t)\rangle=0\end{equation}
\end{subequations}
From these, the ratio
\begin{equation}
\tan\vartheta=-\frac{\langle \sigma_x\rangle_{t,x}}{\langle \sigma_y\rangle_{t,x}}=\frac1{2}\left(\frac{\cosh\frac{t-x}{\alpha}}{\cosh\frac{t+x}{\alpha}}-\frac{\cosh\frac{t+x}{\alpha}}{\cosh\frac{t-x}{\alpha}}\right)
\end{equation}
takes a nice symmetric form containing the ``light cone'' rays $t\pm x$, which govern the speed of propagation of the evolved wavefunction. At long time $t$, however, we observe that $\vartheta$ is still zero at the original location $x=0$ of the TBM state, although the bulk states should have $\gamma =\pm\frac{\pi}{2}$. 

One can also directly obtain the pseudospin expectation amplitude
\begin{equation}
|\langle \boldsymbol\sigma \rangle_{t,x}|=\text{sech}^2\frac{t-x}{\alpha}+\text{sech}^2\frac{t+x}{\alpha}
\label{spinmag}
\end{equation}
which is manifestly a superposition of contributions from two oppositely traveling pulse trains at $x\pm t$. This result (Eq.~\ref{spinmag}) can also be obtained be obtained through Eq.~\ref{sigmaxtmag} with $\hat{\vec d}(p_x)=(p_x,0,0)/|p_x|$. Alternatively, one may also notice that $\hat{\vec d}(p_x)$ is odd, and proceed via Eq.~\ref{odd}. Since $\bold S$ is hence purely imaginary, $\langle \boldsymbol\sigma \rangle_{t,x}$ lies in the $\hat{\bold e}_1$-$\hat{\bold e}_2$ plane. With $\hat{\vec{s}}\propto \hat{\bold e}_2\perp \bold S$, only the first two terms of Eq. \ref{odd} survives, and its evaluation is simple.

\section{Quench dynamics on a lattice}

In following section, we discuss the quench dynamics of the TBM in a lattice. 

%\blue{NEW CONTENT FROM HERE ONWARDS}\\
\begin{figure}
\centering
\subfloat[]{\includegraphics[width=.49 \linewidth]{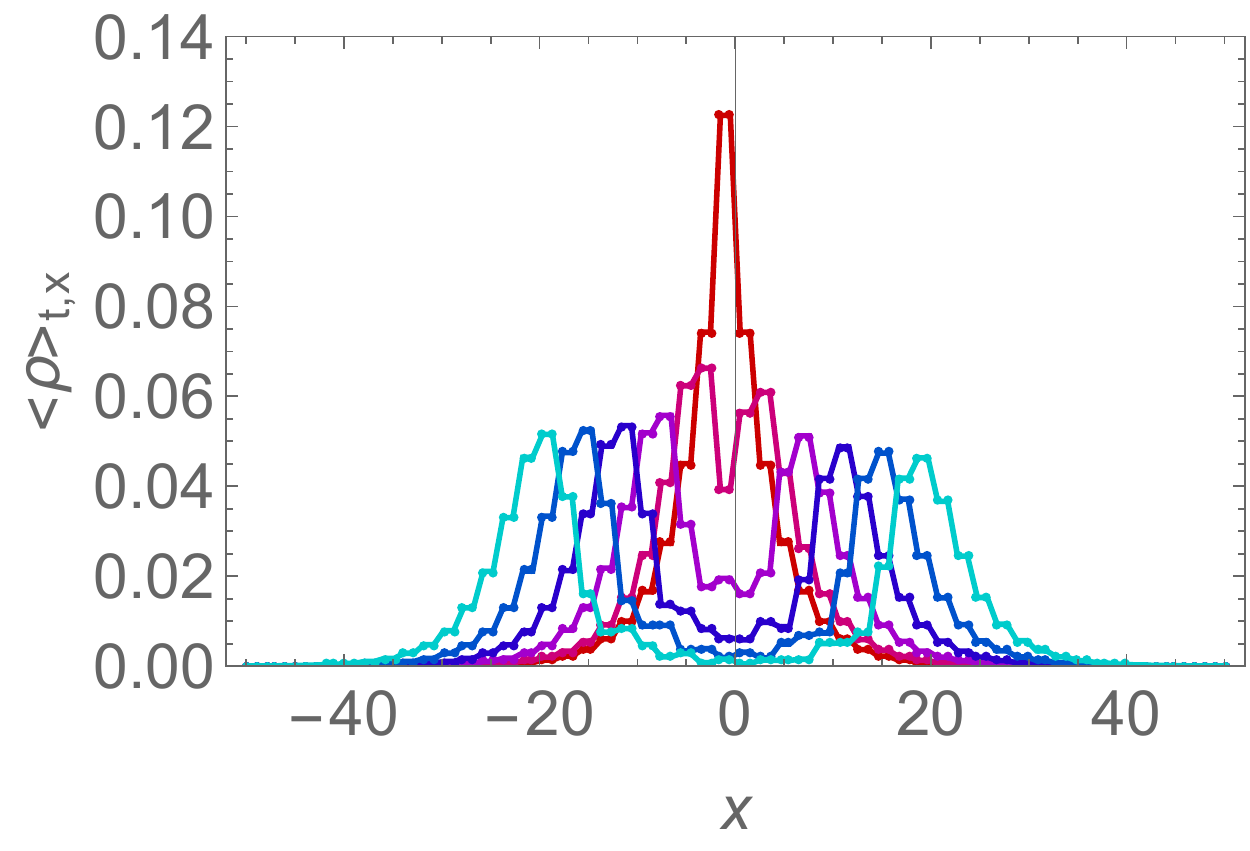}}
\subfloat[]{\includegraphics[width=.49 \linewidth]{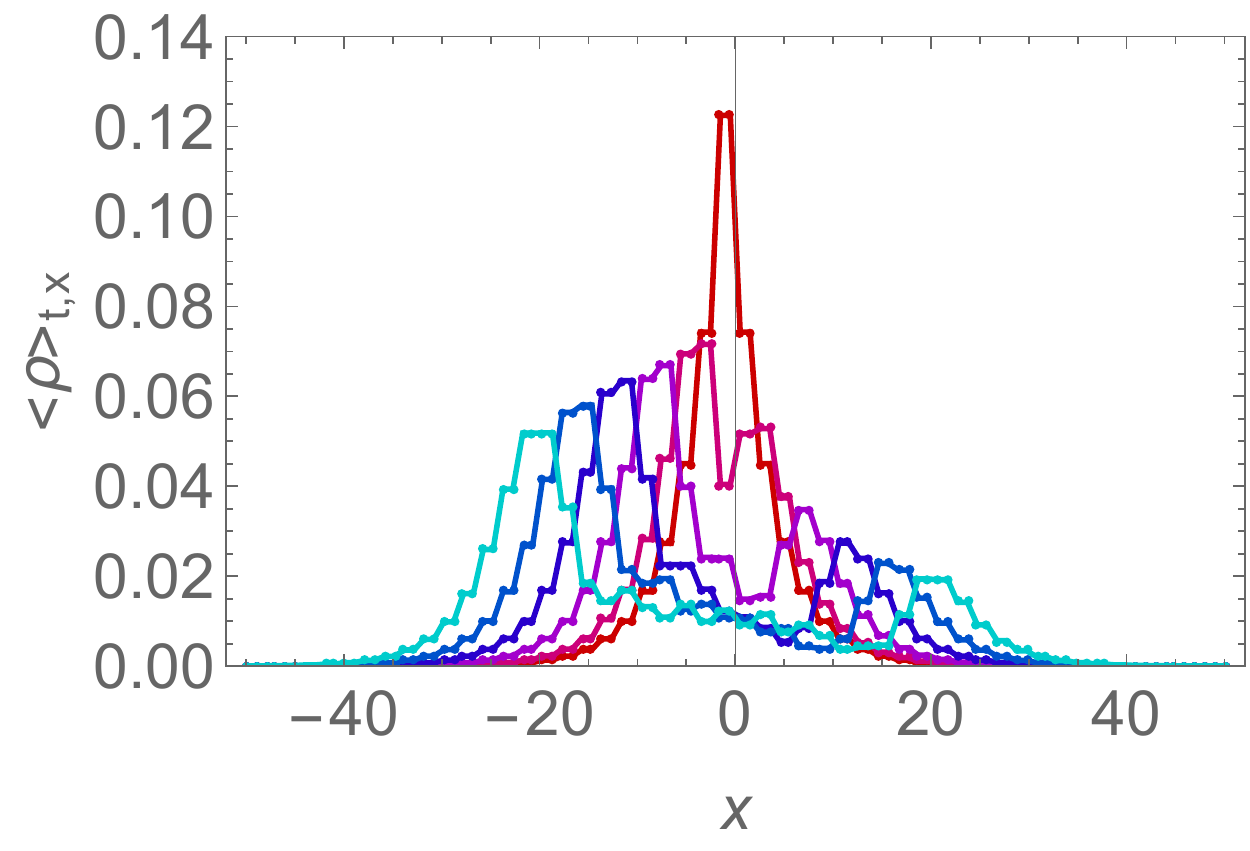}}\\
\subfloat[]{\includegraphics[width=.49 \linewidth]{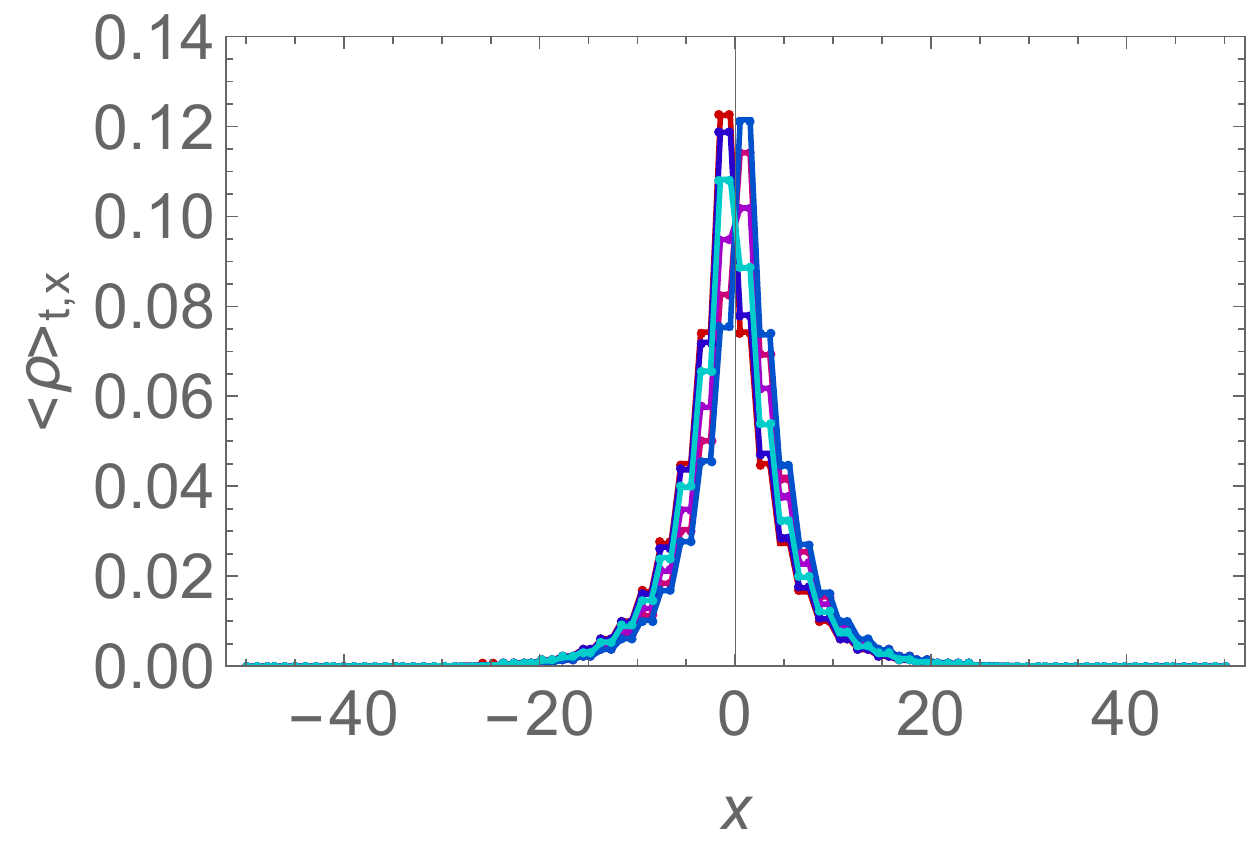}}
\subfloat[]{\includegraphics[width=.49 \linewidth]{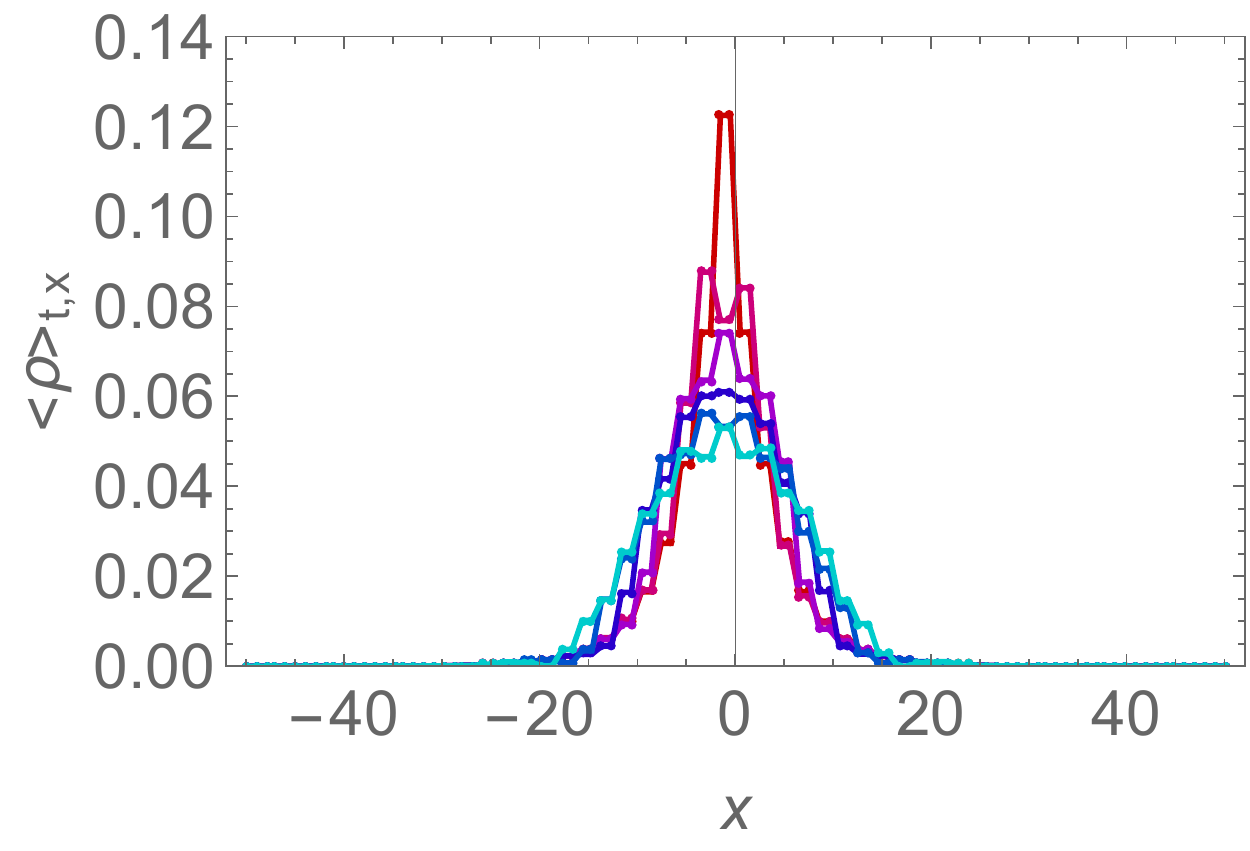}}\\
\subfloat[]{\includegraphics[width=.9 \linewidth]{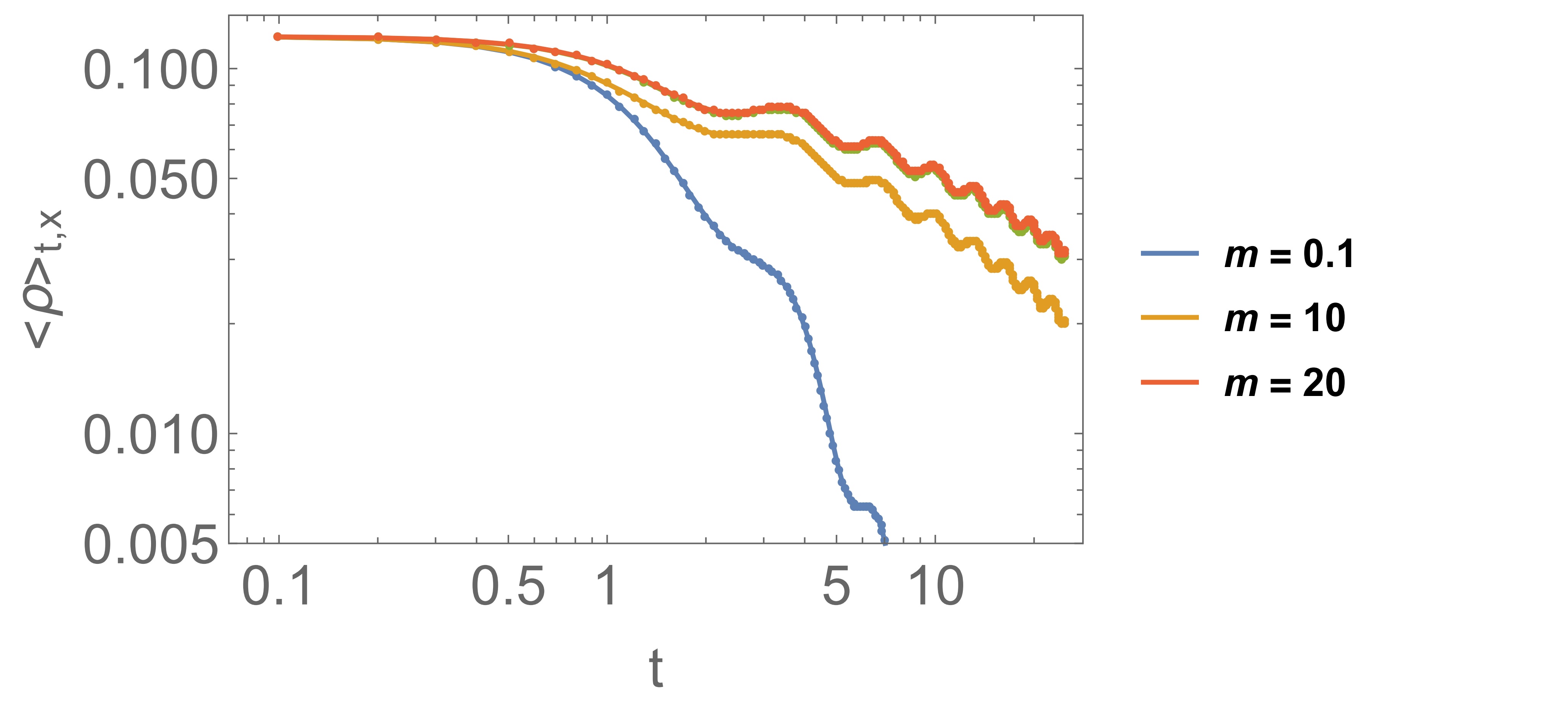}}
\caption{(a)-(d): Post-quench evolution of the initial exponentially localized state, with red to blue representing $t=0,2,4,6,8,10$. Evolution behavior for (a) $m=0.01$, (b) $m=0.1$, (c) $m=-1$ (topological post-quench $H_F$) and (d) large $m=10$. These evolutions agree with Fig.~S1, with slow decay at large $m$. (e) Frozen and melting regimes occur for large $m$ but not small $m$. Positions are given in units of lattice spacing $a$, and times $t$ in units of $v/\alpha$.    
}
\label{plotm}
\end{figure}

\subsection{Dynamics for a localized initial state}

Here we demonstrate that the main conclusions of this work - that localized initial states decay via a frozen followed by a melting regime - still hold on a lattice. This is fortunate, since a continuum model may predict different phenomena, such as the static chiral magnetic effect~\cite{vazifeh}, since many real materials are modelled by tight-binding lattice models. The lattice approach is not expected to give rise to marked differences because the expansion $\epsilon_p=\sqrt{\Delta^2+\hbar^2v^2p_x^2}\approx \Delta + (\hbar vp_y)^2/(2\Delta)$, which underscores the existence of the frozen and melting regimes, is a long wavelength property (we will consider $p_y=0$ since nonzero $p_y$ can be absorbed into $\Delta$). 

%The 
In what follows we use post-quench {\it lattice} Hamiltonian (half-BHZ lattice model) as %is chosen to be 
\begin{eqnarray}
H_F(\bold p)=&&v_F(\sin (p_x a) \sigma_x +\sin (p_y a) \sigma_y)/a\notag\\
&& + (m+2-\cos (p_x a)-\cos (p_y a))\sigma_z\notag\\
\label{HF}
\end{eqnarray}
with a bulk gap $2m$ and topological modes when $-2<m<0$. $a$ is the lattice constant and $v_F$ is the Fermi velocity. Henceforth, we shall work in units of $a=1$ and $v_F=1$ for notational simplicity. The real space lattice consists of two sublattices which are staggered by onsite energies of $\pm (m+2)$. Adjacent unit cells are connected by intra-sublattice hoppings of $\mp \frac1{2}$ in both $x$ and $y$ directions. They are also connected by inter-sublattice hoppings of $\pm \frac{i}{2}$ in these two directions.

From Fig.~\ref{plotm}e, it is evident that the quench on the lattice decays in a similar fashion to the quench in the continuum approach (discussed in the main text), with frozen and then melting regimes that arise at large $m$. At large $m$, similar decay behavior is observed even if the initial state is chosen to be a solution to a hard-wall boundary cutoff (see discussion below). Since a topological in-gap mode also exists for $-2<m<0$, we can also study the post-quench behavior in the presence of a post-quench topological mode as an added bonus. Indeed, there is little post-quench evolution when the post-quench system also possess a boundary mode.

Note that the quench dynamics depend only on the profile of the $t=0$ initial state, and not its origin. In other words, any exponentially localized initial state should have similar evolution behavior, whether it is of topological origin or not. In Fig.~\ref{plotm}a-d, we consider an initial localized state at $x=0$, with an exponential decay length scale chosen to be $7$ lattice sites, a value that ensures a reasonable continuum profile approximation while not requiring a prohibitively large lattice size. The latter is chosen to contain $L=100$ sites, large enough such that Loschmidt echos from boundary reflection do not feature prominently.

\begin{figure}
\centering
\subfloat[]{\includegraphics[width=.49 \linewidth]{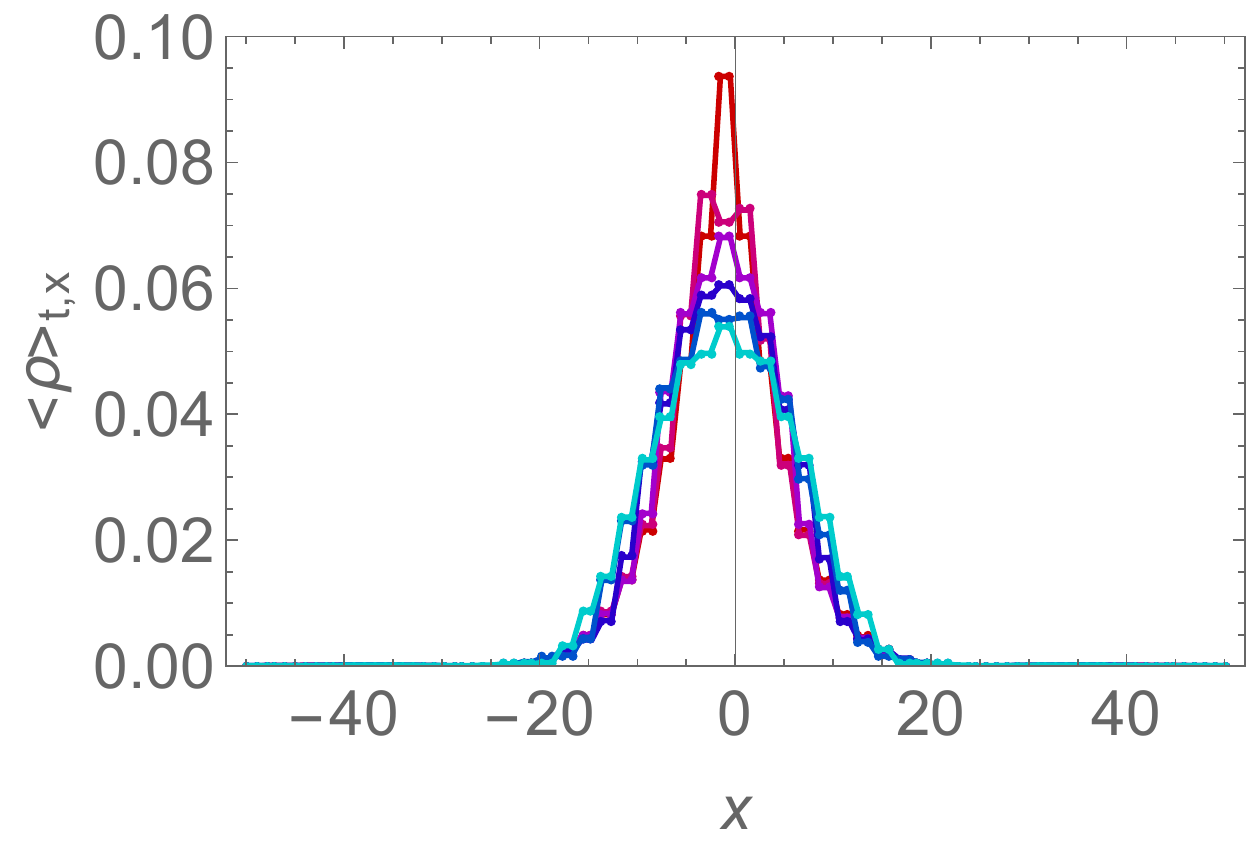}}
\subfloat[]{\includegraphics[width=.49 \linewidth]{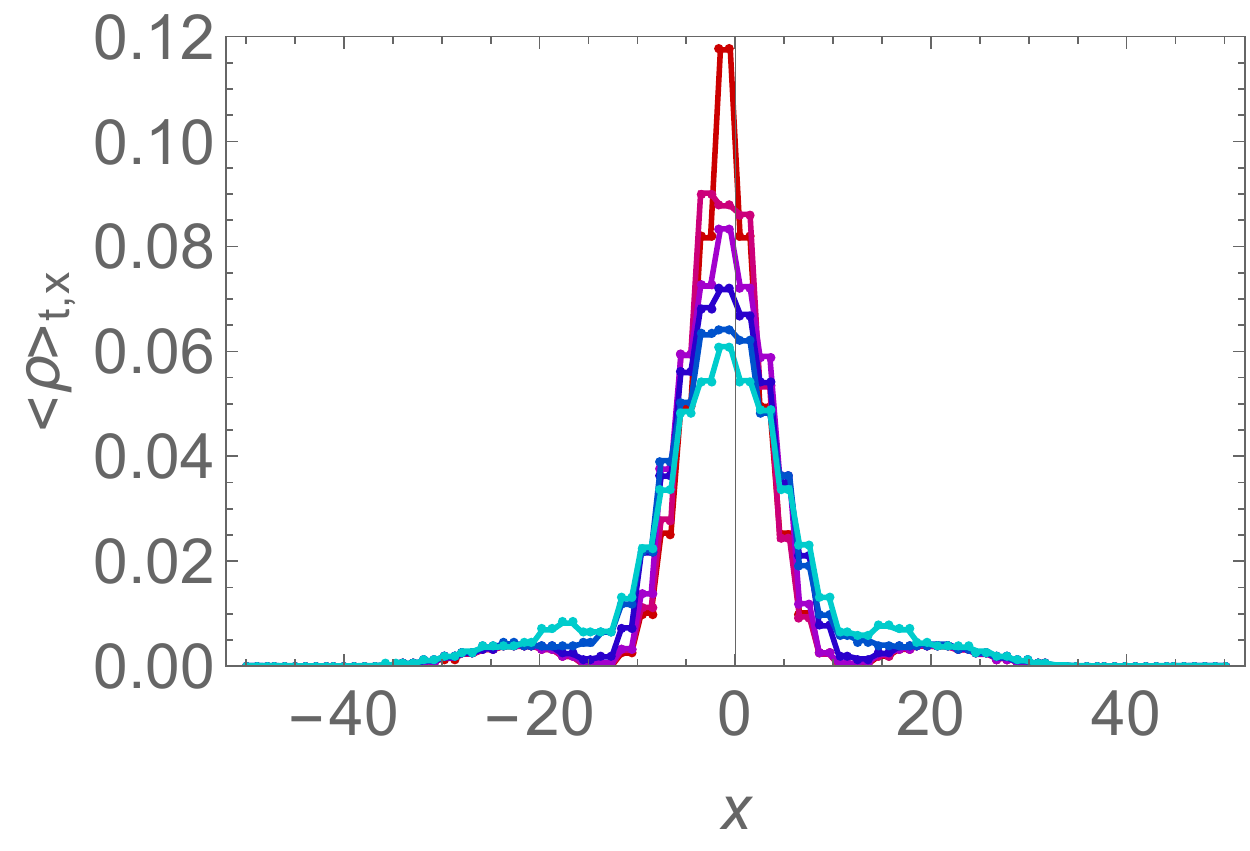}}\\
\subfloat[]{\includegraphics[width=.49 \linewidth]{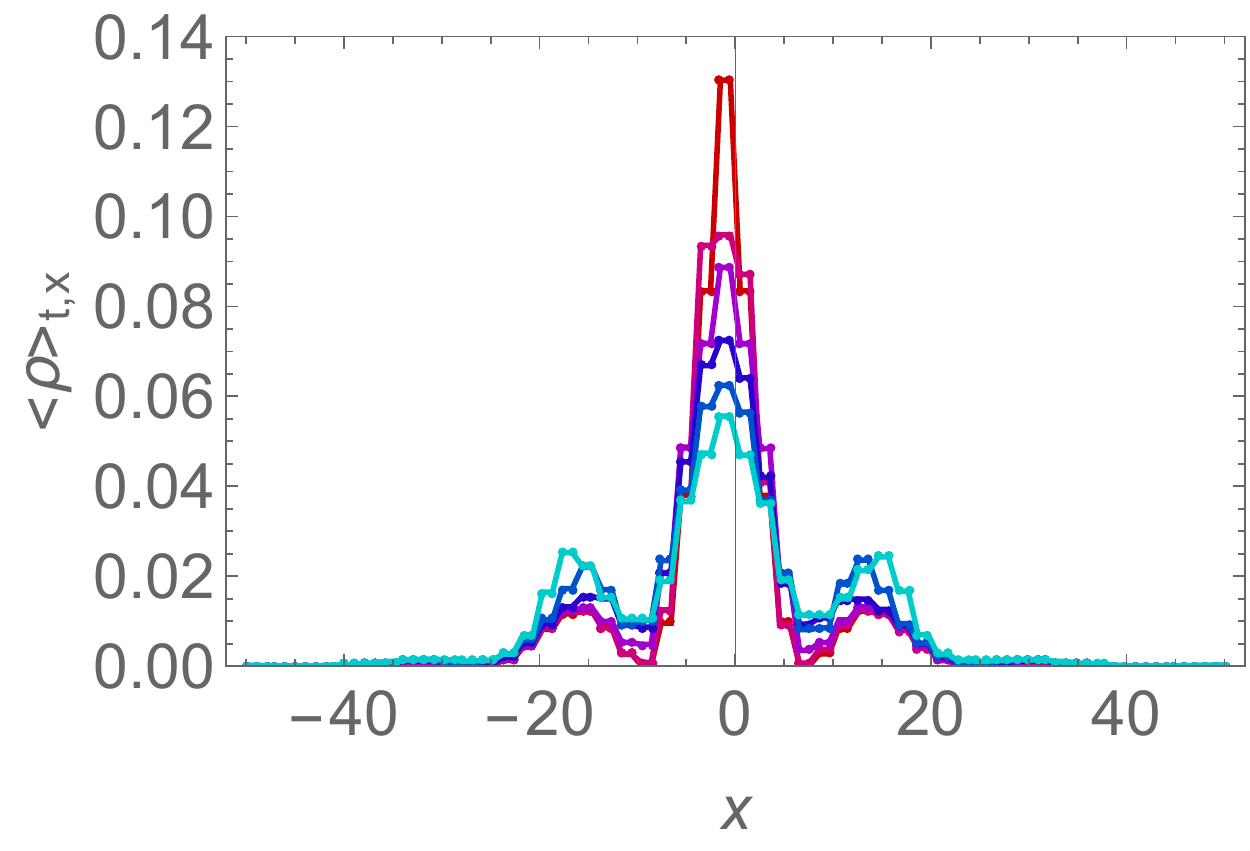}}
\subfloat[]{\includegraphics[width=.49 \linewidth]{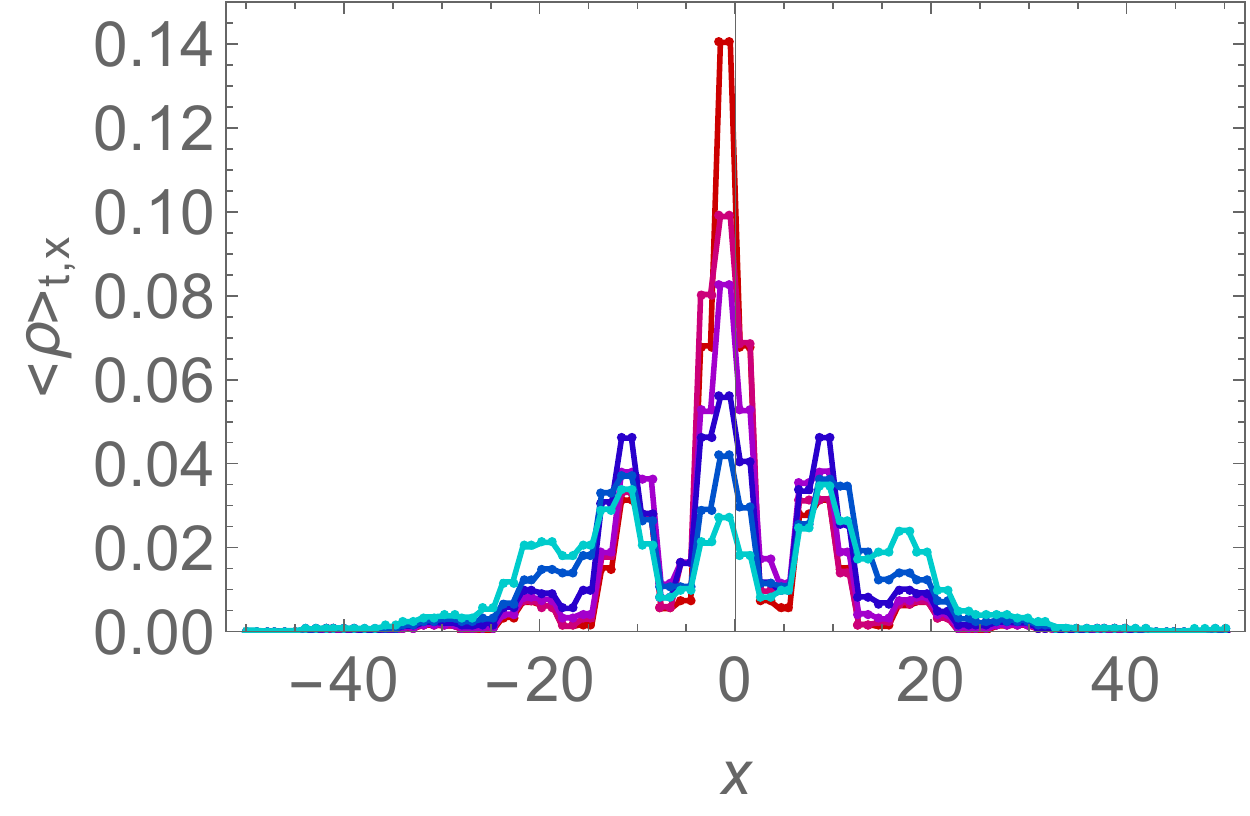}}\\
\subfloat[]{\includegraphics[width=.49 \linewidth]{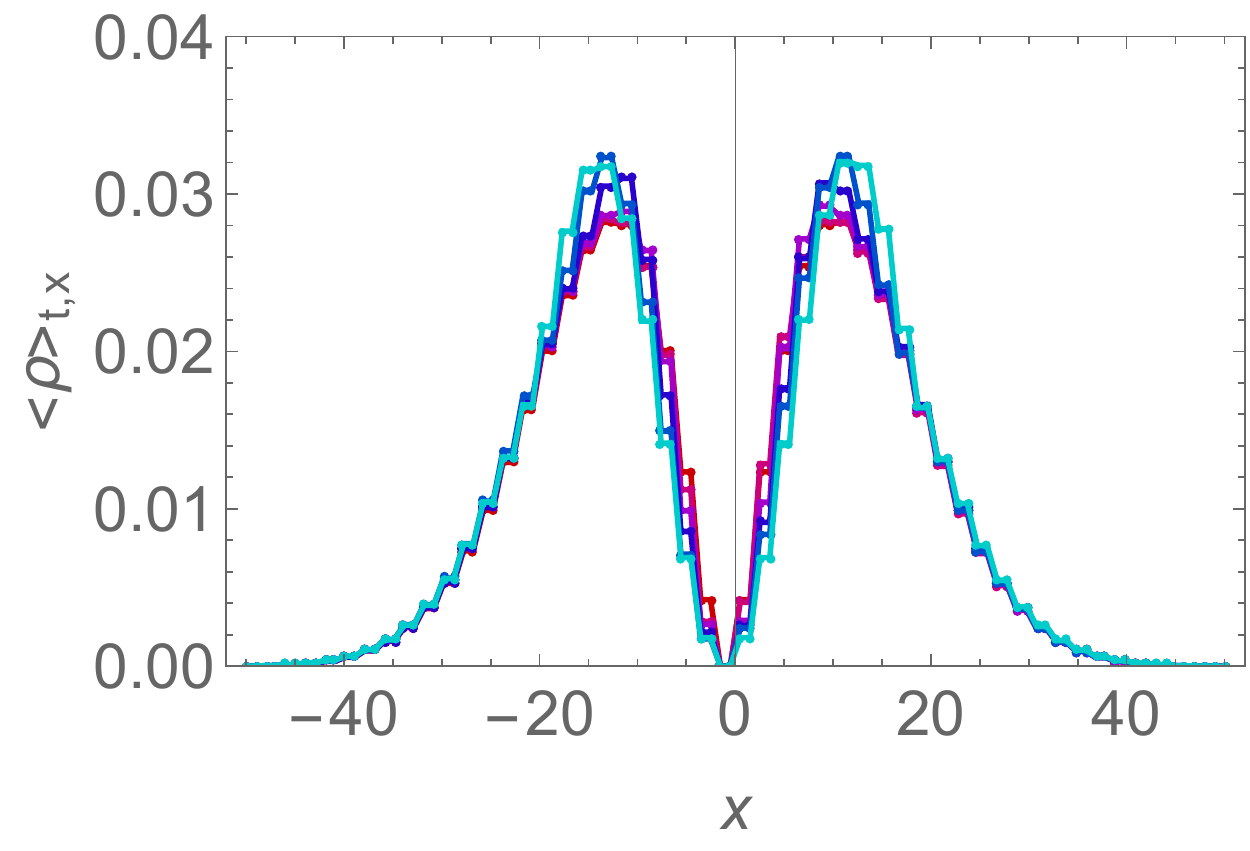}}
\subfloat[]{\includegraphics[width=.49 \linewidth]{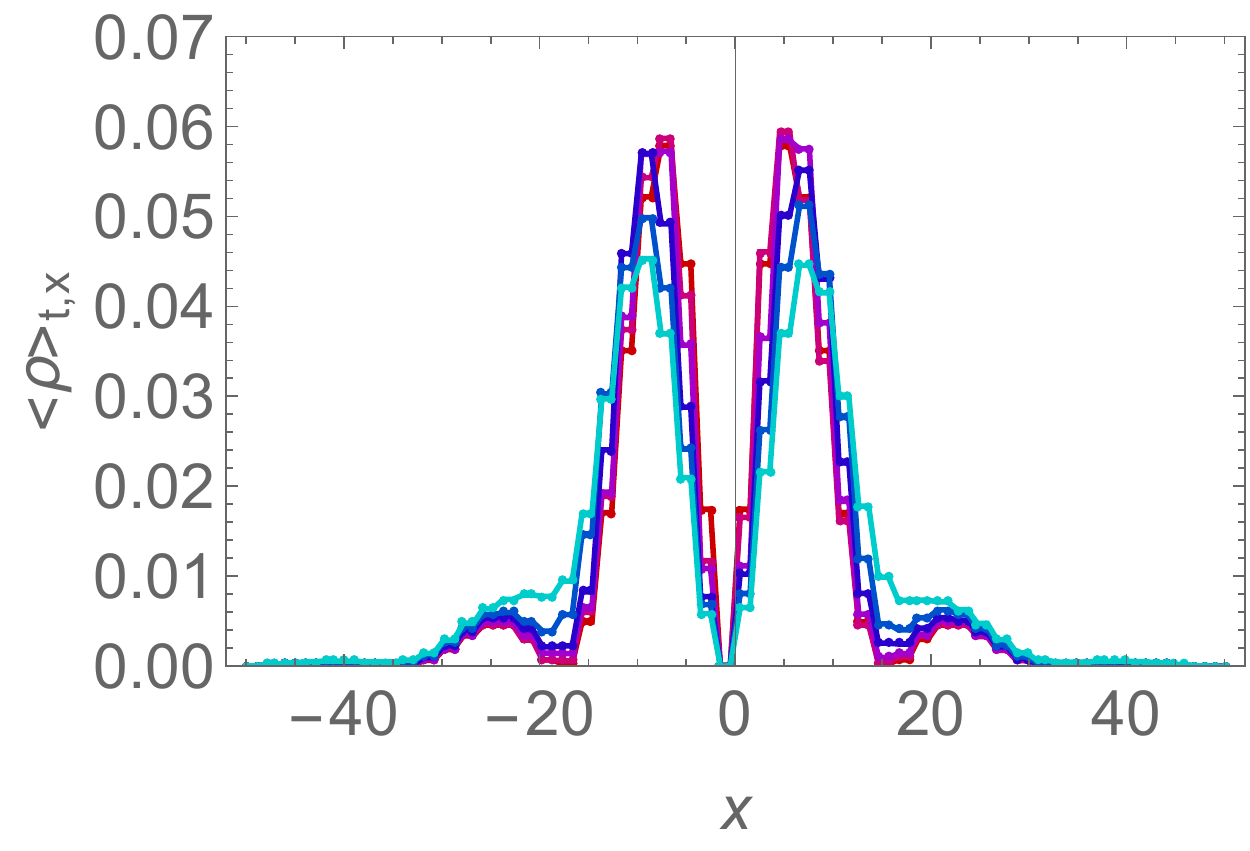}}\\
\subfloat[]{\includegraphics[width=.88 \linewidth]{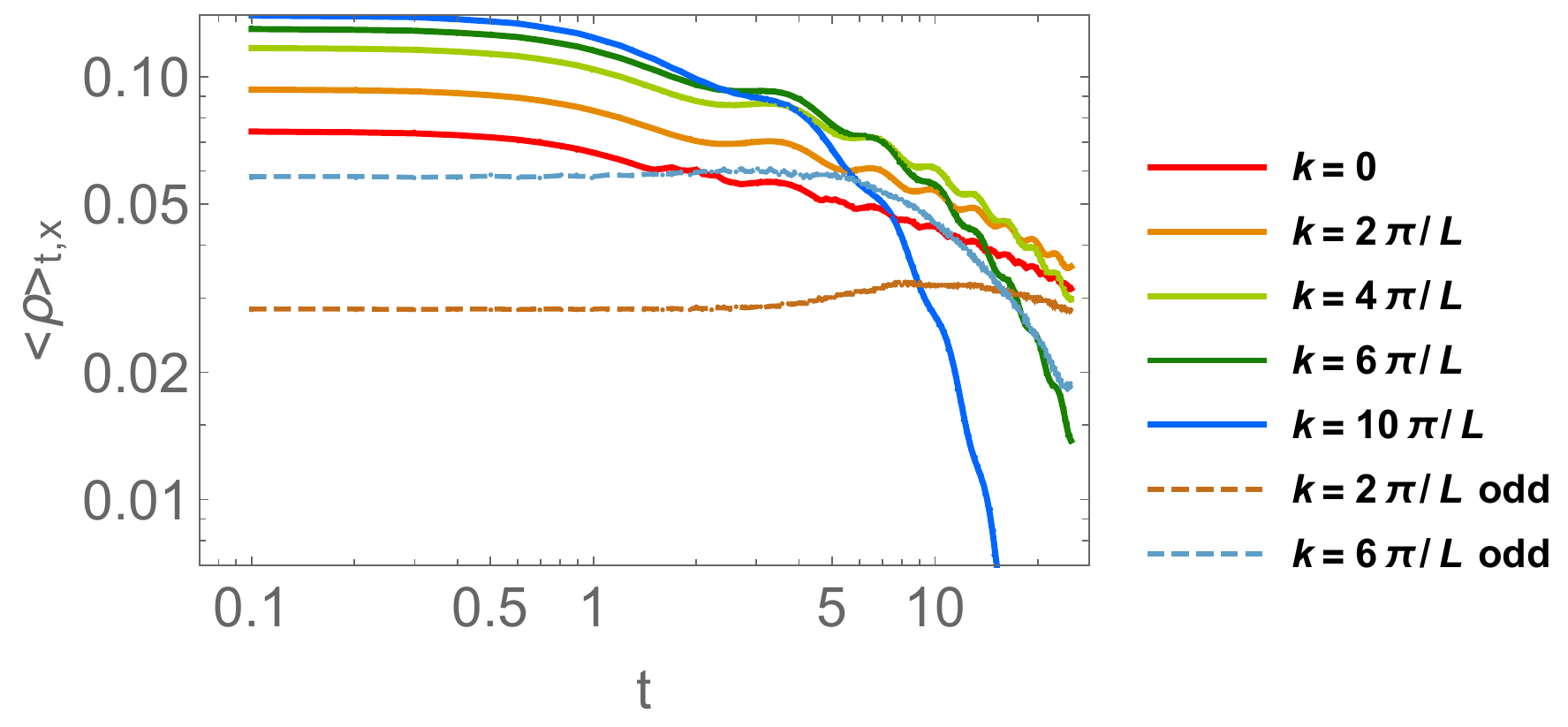}}
\caption{(a-d) Post-quench evolution of the initial exponentially localized state multiplied by a even oscillatory profile with wavevectors $k=1,2,3,5\times 2\pi/L$ respectively, with red to blue representing $t=0,2,4,6,8,10$. (e-f) Analogous evolution with an odd oscillatory profile about $x=0$, with wavevectors $k=1$ and $3\times 2\pi/L$ respectively. (g) Frozen and melting regimes for small $k$ crossovers to frozen followed by exponential decay for larger $k$. Interestingly, odd cases without a central localized part of the wavefunction are essentially frozen for a long time. Positions are given in units of lattice spacing $a$, and times $t$ in units of $v/\alpha$.    }
\label{plotk}
\end{figure}

\subsection{Dynamics for a spatially oscillating initial state}

We now investigate the evolution when the initial state is not tightly localized, but instead possess oscillatory spatial profile characterized by a nonzero wavevector $k$ i.e. $\Psi^{(0)}(x)\sim e^{-\kappa|x|}\cos kx$ for Fig.~\ref{plotk}(a-d) and $\Psi^{(0)}(x)\sim e^{-\kappa|x|}\sin kx$ for Fig.~\ref{plotk}(e-f). From Fig.~\ref{plotk}g, the decay crosses over from power-law to exponential decay as $k$ increases. However, if the spatial oscillation is phase shifted such that the wavefunction no longer has a central peak (Fig.~\ref{plotk}e-f), the decay become even slower.

\subsection{Dynamics for an initial state localized at a hard-wall boundary}

Interestingly, the dynamics for an initial state that is localized at a hard-wall boundary still exhibit slower decay at large post-quench gap $m$, although the frozen and melting regimes are less distinct due to the presence of large momentum sectors arising from sharper gradients in real space.

\begin{figure}
\centering
\subfloat[]{\includegraphics[width=.49 \linewidth]{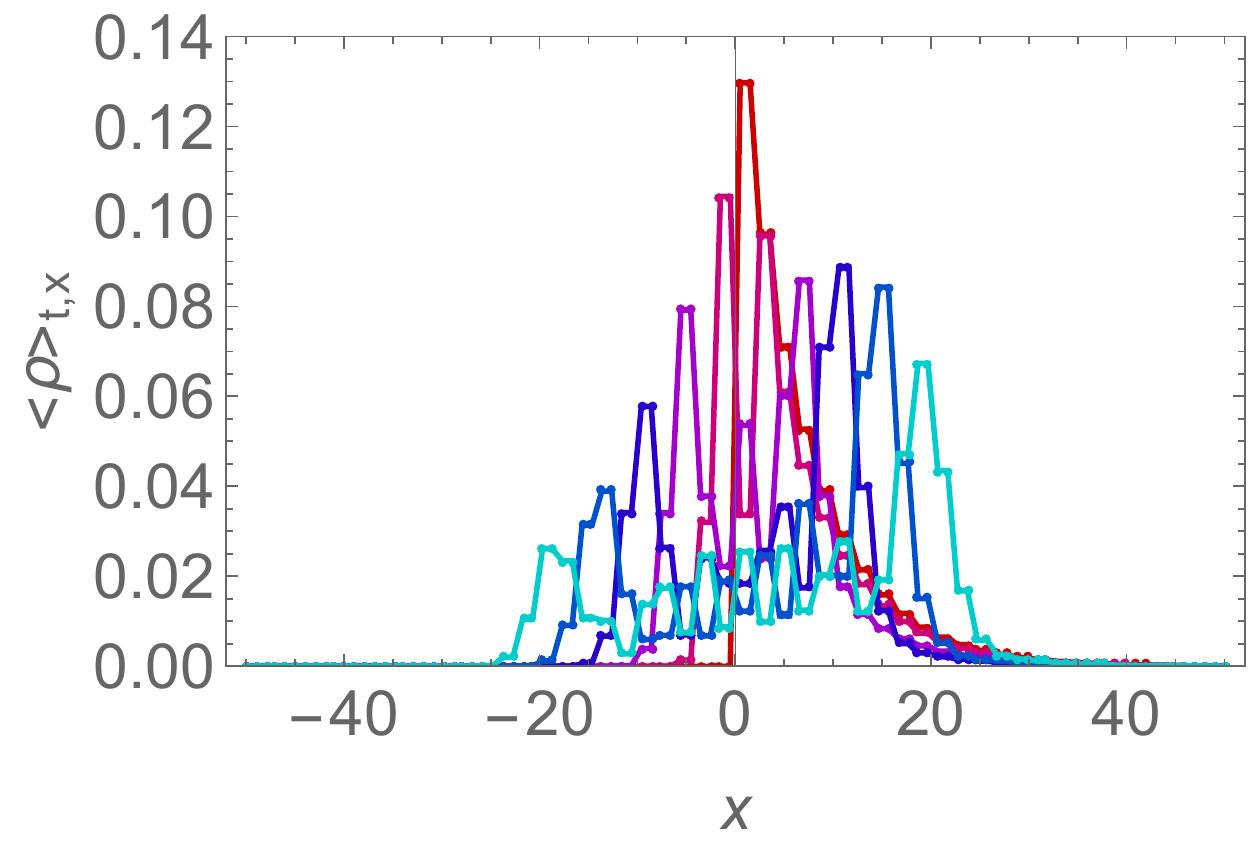}}
\subfloat[]{\includegraphics[width=.49 \linewidth]{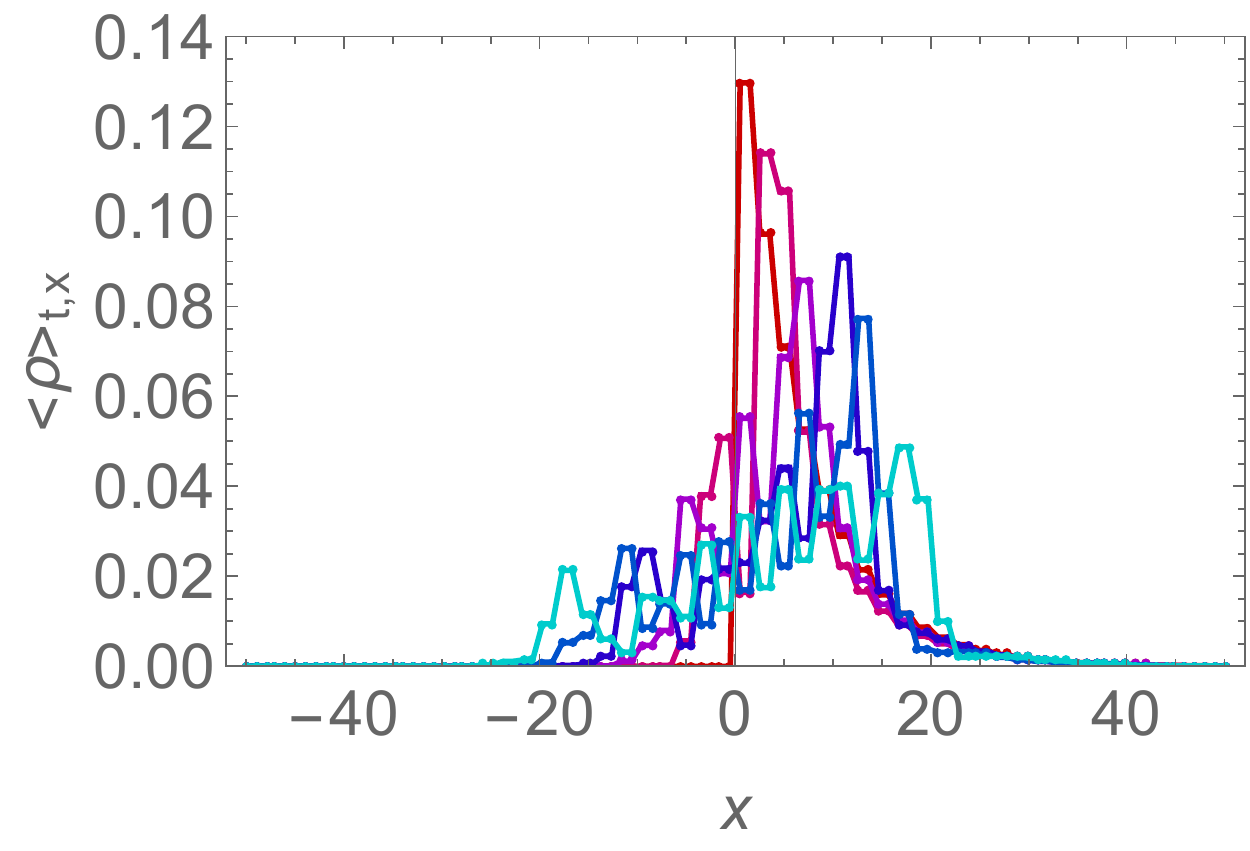}}\\
\subfloat[]{\includegraphics[width=.49 \linewidth]{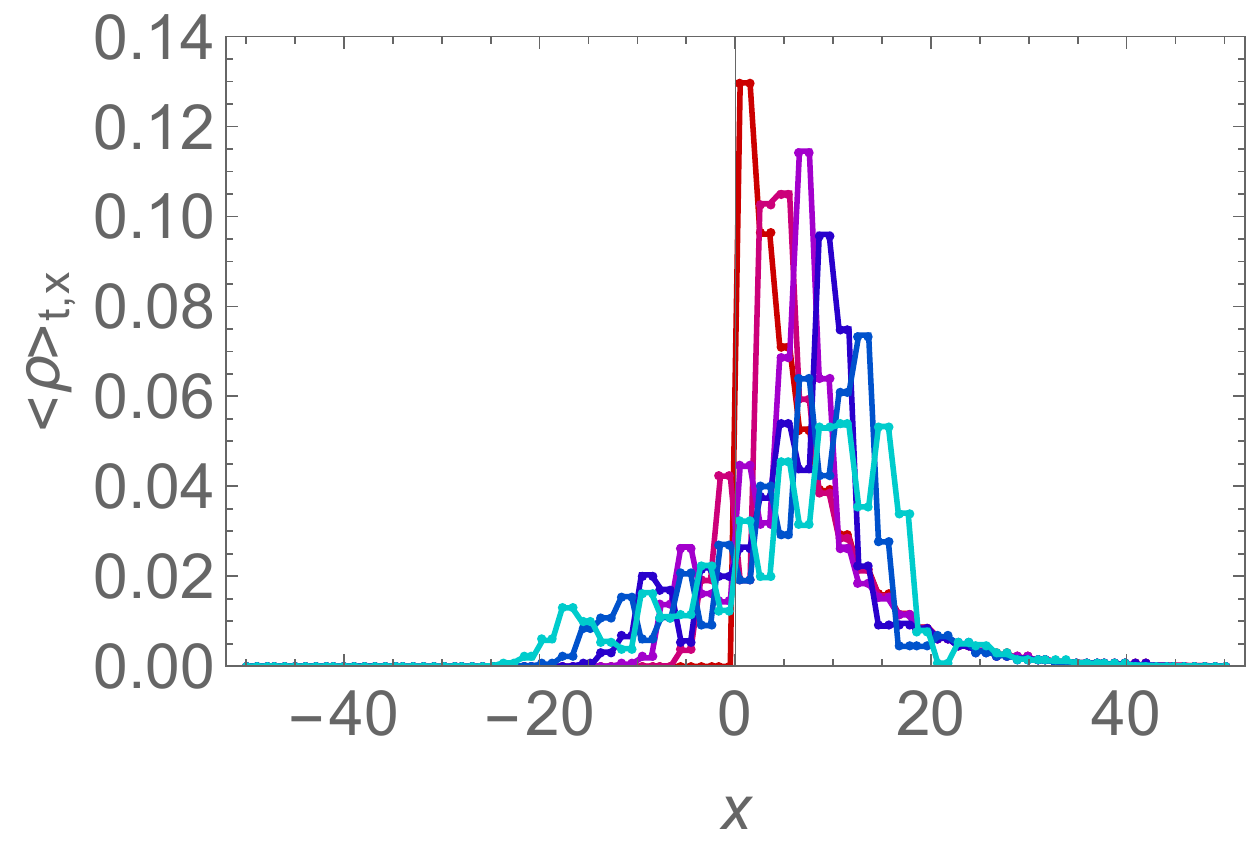}}
\subfloat[]{\includegraphics[width=.49 \linewidth]{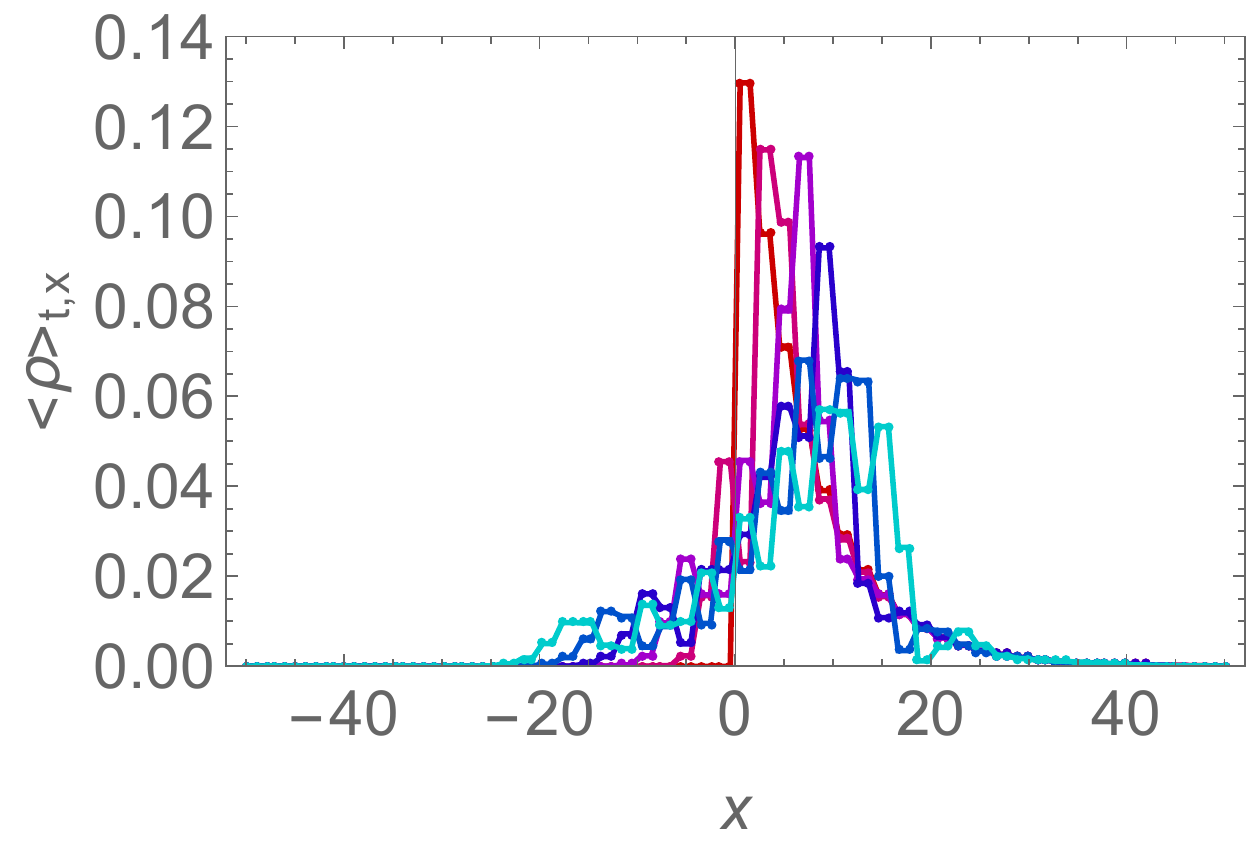}}\\
\subfloat[]{\includegraphics[width=.88 \linewidth]{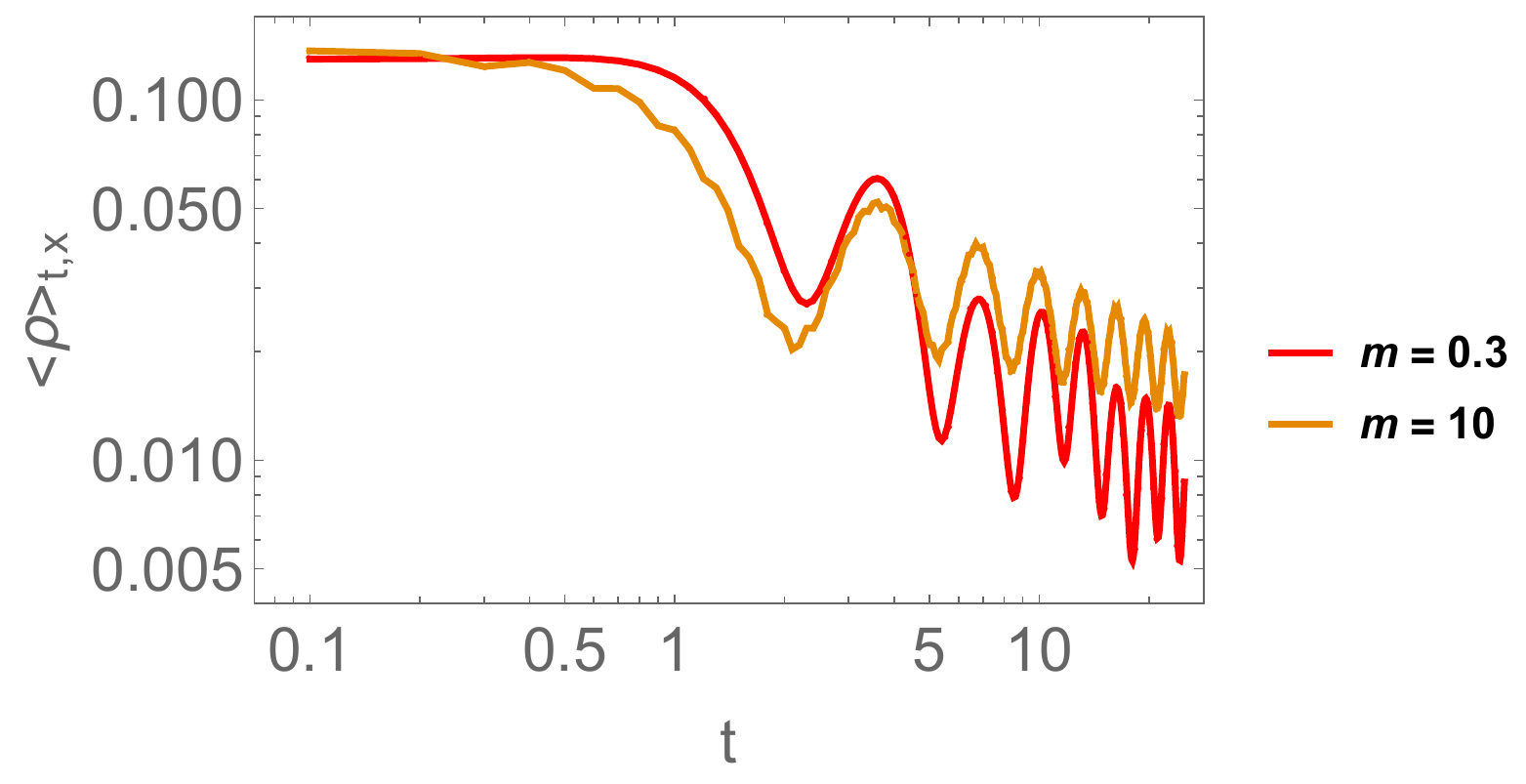}}
\caption{(a-d) Post-quench evolution of an initial state localized at a hard-wall boundary at $x=0$, with red to blue representing $t=0,2,4,6,8,10$. (e) We still observe power-law melting behavior, with large $m$ cases evidently decaying more slowly. The frozen regime is however less evident due to the presence of high-gradient contributions in the initial state. Positions are given in units of lattice spacing $a$, and times $t$ in units of $v/\alpha$.   }
\label{plotOBC}
\end{figure}

A hard-wall boundary that hosts the initial state is implemented by open boundary conditions of our lattice model Eq.~\ref{HF} at $x=0$, with the pre-quench system in the topological regime with $m=-1$. Since the bonds between $x=0$ and $x<0$ sites are physically missing, the the initial wavefunction is abruptly terminated at $x=0$, as given by the red curves in Fig.~\ref{plotOBC}(a-d). They subsequently decay in a qualitatively similar fashion as in an exponentially localized initial state (Fig.~\ref{plotm}), though somewhat less cleanly. This is because the hard-wall cutoff introduces higher momentum harmonics, which do not strictly decay in a power-law fashion as explained in the previous subsection on the spatially oscillating initial state. Even so, because the initial state here is not dominated by a nonzero wavevector, the state decay is still power-law, albeit less cleanly so.

\section{Post-quench survival probability}

\begin{figure*}
\centering
\subfloat[]{\includegraphics[width=.33 \linewidth]{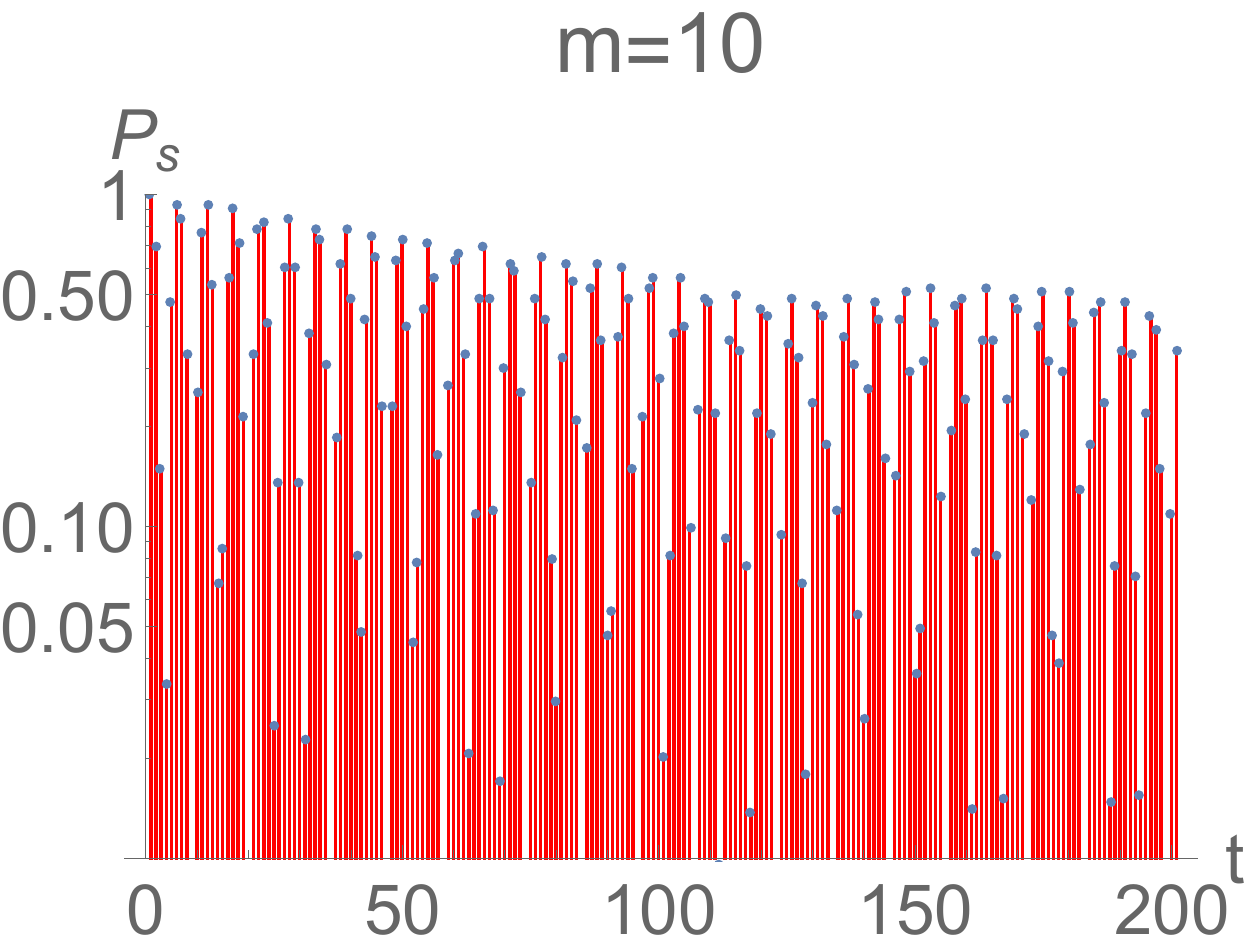}}
\subfloat[]{\includegraphics[width=.33 \linewidth]{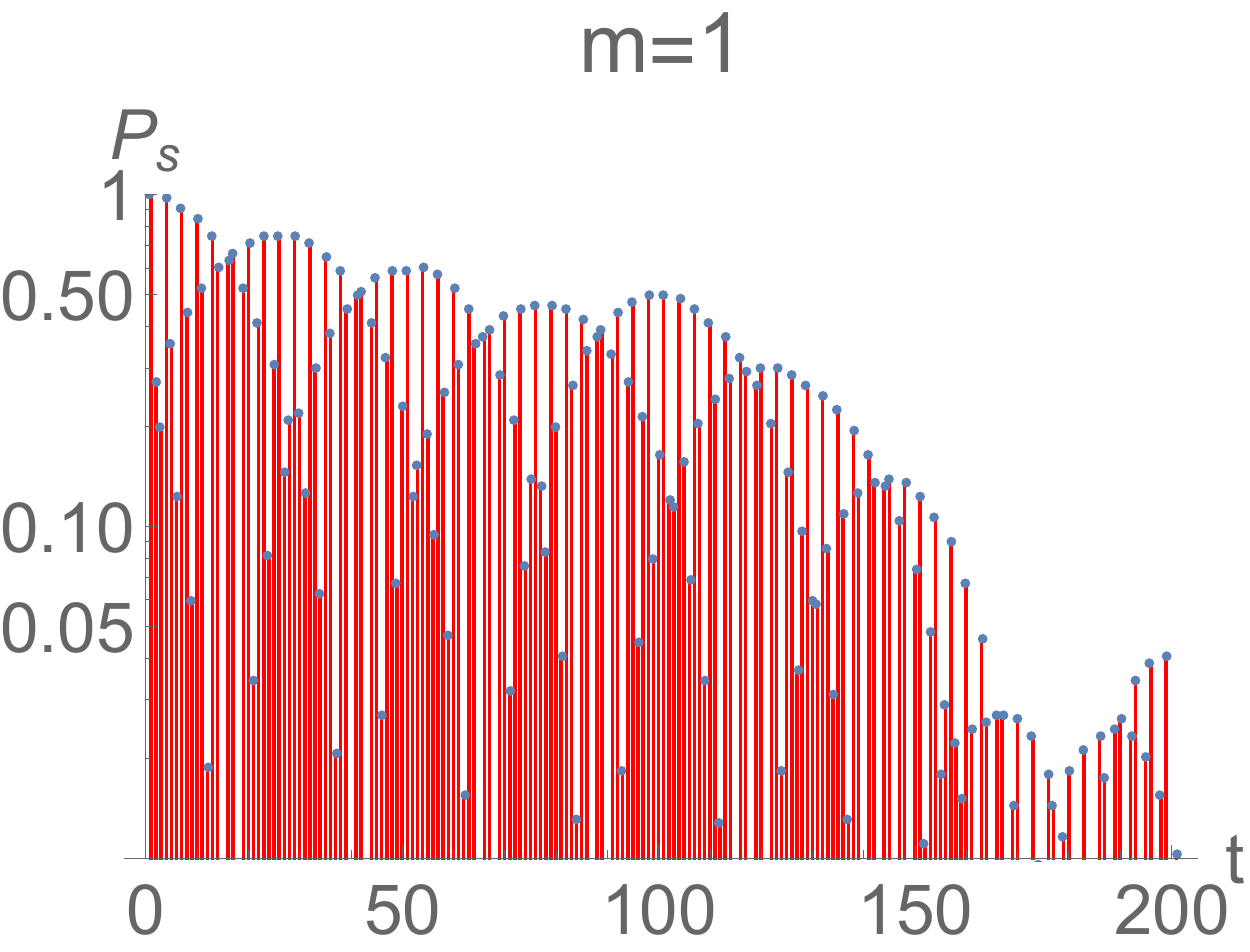}}
\subfloat[]{\includegraphics[width=.33\linewidth]{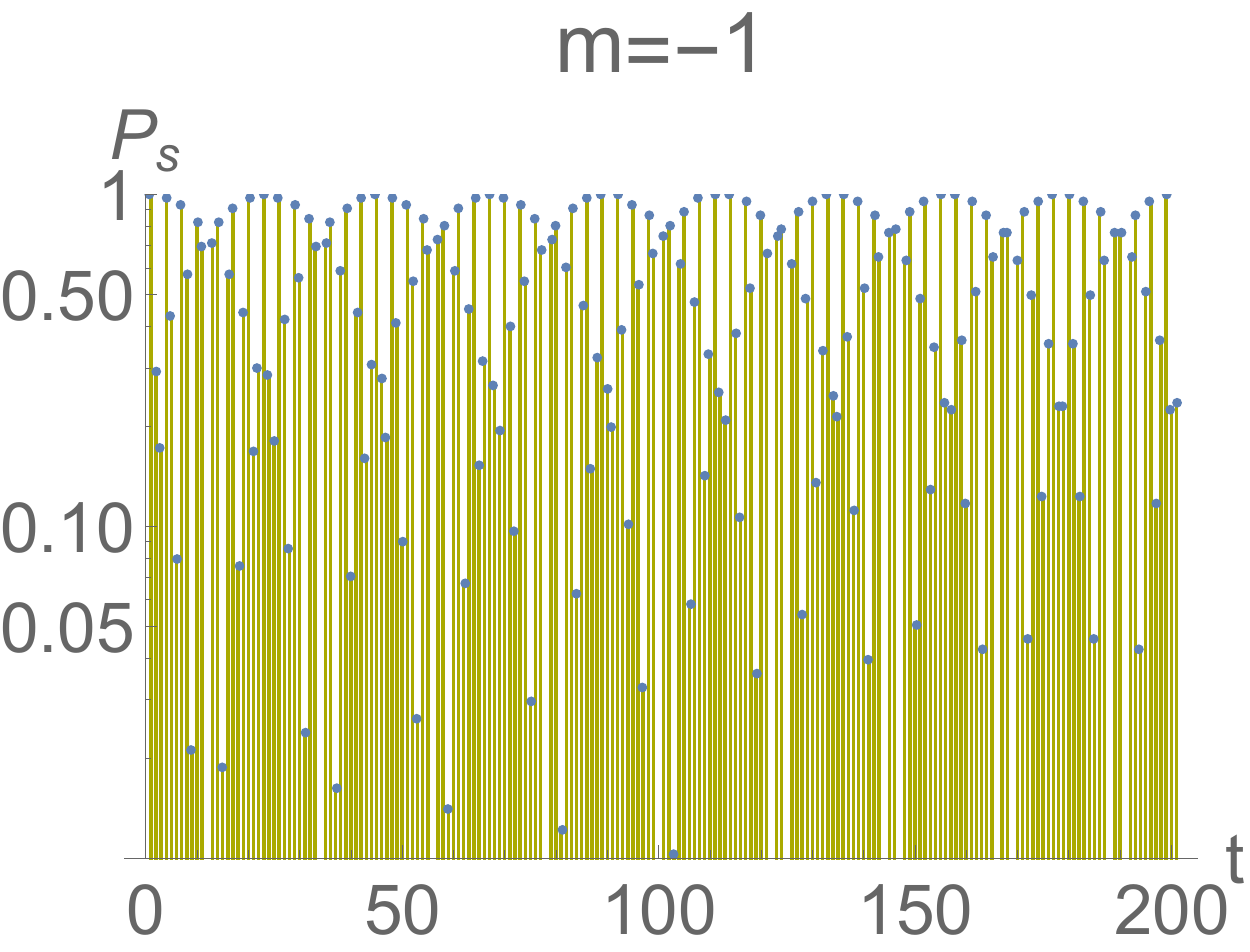}}
\caption{The evolution of survival probability for the initial state of Fig.~\ref{plotm}, due to a quench given by Eq.~\ref{HF}, with time $t$ in units of $v/\alpha$. There is exponential, albeit slow, decay in the large $m$ case, and no long-term decay in the $m=-1$ topological case. The small gap ($m=1$) case decays the fastest.  
}
\label{plotPs}
\end{figure*}

It is interesting to contrast the post-quench behavior of the probability density or pseudospin expectation with the survival probability, which is defined as~\cite{dutta2015,tanay2019}
\begin{equation}
P_s=\left|\sum_n|\langle \Psi^{(0)}|\Psi(t)\rangle|^2e^{-iE_nt}\right|^2
\end{equation}
where $\Psi^{(0)}$ and $\Psi(t)$ are respectively the pre-quench state and the post-quench state, and $E_n$ the post-quench eigenenergies. $P_s$ is a measure of how much the pre-quench and post-quench states overlap. 

From Eq.~\ref{plotPs}, the evolution of $P_s$ does not agree with that of the density. This is understandable since the survival probability concerns the overlap between the initial and evolved state, which is extended and moving in real-space, while the probability density concerns the evolution at a single real-space site. Hence it is not surprising that they are different.

Incidentally, if we look at the survival probability when $m=-1$, when the post-quench Hamiltonian also has a topological edge mode, the survival probability remain close to 1.

\end{document}